%% file: main.tex
\documentclass[sigconf, nonacm]{acmart}

\usepackage{subcaption}
\usepackage{listings}
\usepackage{dsfont}
\usepackage{etoolbox,graphicx,setspace,listings,multicol,xspace,enumitem,booktabs,xcolor}
\usepackage{algorithm}
\usepackage[noend]{algpseudocode}
\usepackage[a-2b]{pdfx}

\newcommand\vldbdoi{10.14778/3611479.3611496}
\newcommand\vldbpages{2897 - 2910}
\newcommand\vldbvolume{16}
\newcommand\vldbissue{11}
\newcommand\vldbyear{2023}
\newcommand\vldbauthors{\authors}
\newcommand\vldbtitle{\shorttitle} 
\newcommand\vldbavailabilityurl{https://github.com/stanford-futuredata/InQuest}
\newcommand\vldbpagestyle{empty} 

\newcommand{\algname}[0]{\text{InQuest}}

\newtheorem{proposition}{Proposition}
\newtheorem{theorem}{Theorem}

\newcommand*\LSTfont{\Small\ttfamily\SetTracking{encoding=*}{-60}\lsstyle}
\newcommand{\minihead}[1]{{\vspace{.45em}\noindent\textbf{#1.} }}

\begin{document}
\title{Accelerating Aggregation Queries on Unstructured Streams of Data}

\author{Matthew Russo}
\affiliation{%
  \institution{Stanford University}
}
\email{russom@stanford.edu}

\author{Tatsunori Hashimoto}
\affiliation{%
  \institution{Stanford University}
}
\email{thashim@stanford.edu}

\author{Daniel Kang}
\affiliation{%
  \institution{University of Illinois Urbana-Champaign}
}
\email{ddkang@illinois.edu}

\author{Yi Sun}
\affiliation{%
  \institution{University of Chicago}
}
\email{yi.sun@uchicago.edu}

\author{Matei Zaharia}
\affiliation{%
  \institution{Stanford University}
}
\email{matei@cs.stanford.edu}

\input{tex/abstract}

\maketitle

\pagestyle{\vldbpagestyle}
\begingroup\small\noindent\raggedright\textbf{PVLDB Reference Format:}\\
\vldbauthors. \vldbtitle. PVLDB, \vldbvolume(\vldbissue): \vldbpages, \vldbyear.\\
\href{https://doi.org/\vldbdoi}{doi:\vldbdoi}
\endgroup
\begingroup
\renewcommand\thefootnote{}\footnote{\noindent
This work is licensed under the Creative Commons BY-NC-ND 4.0 International License. Visit \url{https://creativecommons.org/licenses/by-nc-nd/4.0/} to view a copy of this license. For any use beyond those covered by this license, obtain permission by emailing \href{mailto:info@vldb.org}{info@vldb.org}. Copyright is held by the owner/author(s). Publication rights licensed to the VLDB Endowment. \\
\raggedright Proceedings of the VLDB Endowment, Vol. \vldbvolume, No. \vldbissue\ %
ISSN 2150-8097. \\
\href{https://doi.org/\vldbdoi}{doi:\vldbdoi} \\
}\addtocounter{footnote}{-1}\endgroup

\ifdefempty{\vldbavailabilityurl}{}{
\vspace{.3cm}
\begingroup\small\noindent\raggedright\textbf{PVLDB Artifact Availability:}\\
The source code, data, and/or other artifacts have been made available at \url{\vldbavailabilityurl}.
\endgroup
}
\input{tex/intro}

\input{tex/overview}
\input{tex/algorithm}

\input{tex/analysis}

\input{tex/evaluation}
\input{tex/rel_work}
\input{tex/conclusion}

\begin{acks}
This research was supported in part by affiliate members and other supporters of the Stanford DAWN project—Ant Financial, Facebook, Google, and VMware—as well as Toyota Research Institute, Cisco, SAP, and the NSF under CAREER grant CNS-1651570. This work is also supported in part by the Open Philanthropy project. Any opinions, findings, and conclusions or recommendations expressed in this material are those of the authors and do not necessarily reflect the views of the National Science Foundation. Toyota Research Institute (“TRI”) provided funds to assist the authors with their research but this article solely reflects the opinions and conclusions of its authors and not TRI or any other Toyota entity.
\end{acks}


\bibliographystyle{ACM-Reference-Format}
\bibliography{sample}

\end{document}

%% file: tex/abstract.tex
\begin{abstract}
Analysts and scientists are interested in querying streams of video, audio, and text to extract quantitative insights. For example, an urban planner may wish to measure congestion by querying the live feed from a traffic camera. Prior work has used deep neural networks (DNNs) to answer such queries in the batch setting. However, much of this work is not suited for the streaming setting because it requires access to the entire dataset before a query can be submitted or is specific to video. Thus, to the best of our knowledge, no prior work addresses the problem of efficiently answering queries over multiple modalities of streams.



In this work we propose \algname{}, a system for accelerating aggregation queries on unstructured streams of data with statistical guarantees on query accuracy. \algname{} leverages inexpensive approximation models (``proxies”) and sampling techniques to limit the execution of an expensive high-precision model (an ``oracle”) to a subset of the stream. It then uses the oracle predictions to compute an approximate query answer in real-time. We theoretically analyzed \algname{} and show that the expected error of its query estimates converges on stationary streams at a rate inversely proportional to the oracle budget. We evaluated our algorithm on six real-world video and text datasets and show that \algname{} achieves the same root mean squared error (RMSE) as two streaming baselines with up to 5.0x fewer oracle invocations. We further show that \algname{} can achieve up to 1.9x lower RMSE at a fixed number of oracle invocations than a state-of-the-art batch setting algorithm.

\end{abstract}

%% file: tex/intro.tex
\begin{figure}[t!]
  \centering
  \includegraphics[width=\linewidth]{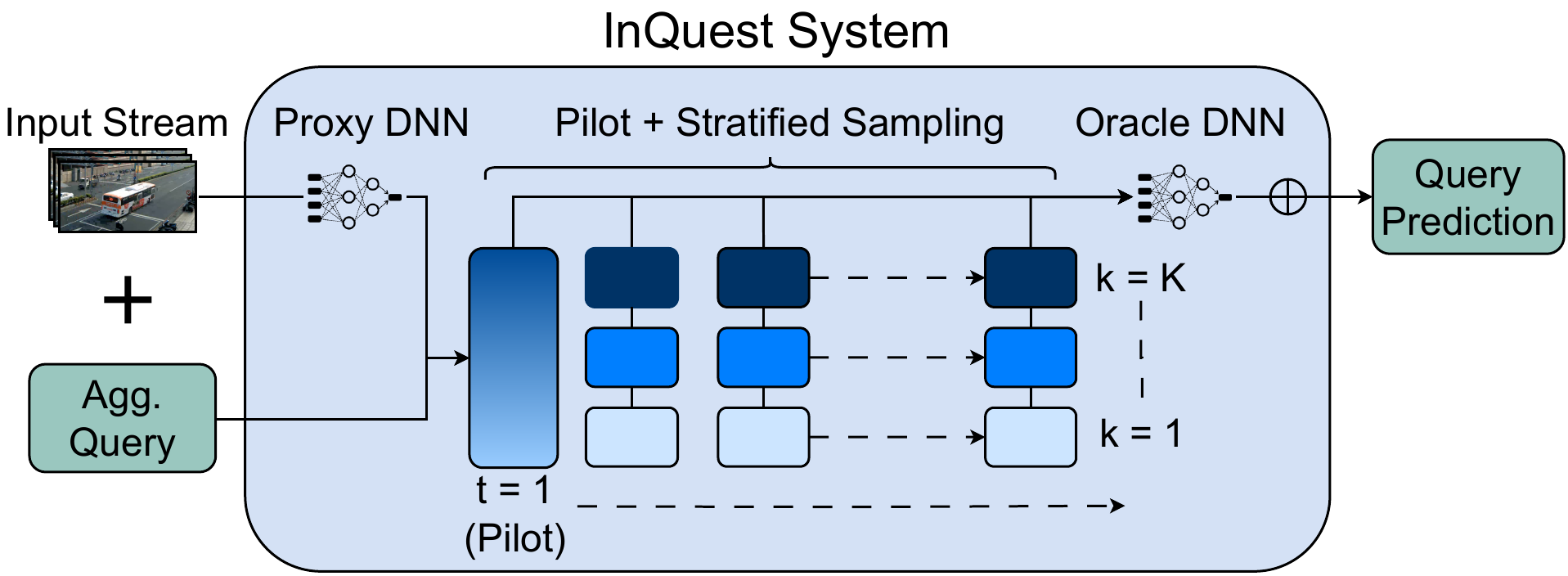}
  \caption{System diagram for \algname{}. The user provides an input stream (e.g., a video stream), an aggregation query, a proxy model, and an oracle. \algname{} processes the stream with the proxy and performs pilot sampling to compute $K$ initial strata. \algname{} applies the oracle to samples drawn from each strata and computes the query prediction.} 
  \label{fig:system-diagram}
\end{figure}

\section{Introduction}
Unstructured streams of data, i.e., streams of data without a well-defined schema (video, audio, and text), are increasingly prevalent. In November 2022, the streaming platform Twitch averaged more than 2 million hours of live video per day \cite{twitch2022data}, while microblogging platforms like Twitter can process 500 million tweets daily \cite{twitter2013data}. Furthermore, streams are increasingly being used to address real-world problems. As an example, a consortium of universities working with the U.S. Forest Service has deployed hundreds of cameras in the western United States to locate wildfires quickly \cite{alertwildfire2022cameras}.


Analysts would like to be able to query these streams of data to extract quantitative insights at minimal cost. For example, a social scientist may wish to quantify the sentiment of a Twitter feed during a presidential debate. The researcher may also want to filter for the subset of tweets mentioning a specific candidate. 

Prior work has leveraged deep neural networks (DNNs) to execute queries over large datasets of unstructured data \cite{hong2020analyzing, poms2018scanner, zhang2017live, fu2019rekall, zhang2020panorama, anderson2019tahoma, hsieh2018focus, kang2018blazeit, kang2017noscope, kang2020jointly}. For example, the social scientist might use a BERT model \cite{devlin2018bert} to execute the following query:
\begin{lstlisting}[
  frame=tb,
  basicstyle=\LSTfont,
  keywordstyle=\color{blue},
  language=SQL,
  aboveskip=4pt,
  belowskip=4pt,
]
SELECT COUNT(positive(tweet)) FROM twitter
WHERE mentions_candidate(tweet)
\end{lstlisting}
The DNN would determine whether the tweet satisfied the query predicate---in this case by mentioning the candidate. It would also compute the statistic of interest, i.e., the sentiment of the tweet. 

One limitation of using DNNs for query processing is that executing large DNNs exhaustively over most real-world datasets can be prohibitively expensive. As a result, recent work has focused on accelerating DNN-based queries \cite{anderson2019tahoma, hsieh2018focus, kang2018blazeit, kang2017noscope, kang2020approximate, kang2021abae, moll2022exsample}. One common approach involves filtering (or sub-sampling) records with inexpensive approximation models (``proxies”) and then applying high-precision models (``oracles”) to extract statistics of interest.




Unfortunately, many of these systems are designed to work in the batch setting and cannot easily be adapted to answer queries over streams. For example, NoScope \cite{kang2017noscope} and Tahoma \cite{anderson2019tahoma} would need to buffer the entire stream in order to train and validate specialized DNNs and model cascades before answering a query. This limits their ability to answer queries over streams in real-time. Other systems can operate in the streaming setting but are limited to processing a specific modality of data such as video \cite{koudas2020video}. Thus, to the best of our knowledge, there is currently no system for efficiently processing aggregation queries over multiple modalities of large unstructured streams of data.





In this work we propose and analyze \algname{}, a system to accelerate aggregation queries over unstructured streams of data with statistical guarantees on query accuracy. By design, \algname{} uses stratified sampling \cite{parsons2014stratified} to (1) compute precise query estimates with better sample efficiency than uniform sampling and (2) provide standard frequentist bounds for our query estimates. \algname{} takes a stream, an aggregation query, a proxy, and an oracle as input. The user also specifies a tumbling window \cite{flink2023syntax} (i.e., a ``segment” length) as well as an oracle budget per segment. As illustrated in \autoref{fig:system-diagram}, \algname{} processes the stream segment-by-segment while producing a query estimate in real-time. For each segment, \algname{} separates records in the stream into disjoint strata based on their proxy estimates. It then runs the oracle on records sampled from each stratum. Finally, \algname{} computes an estimate of the query answer based on the oracle’s predictions on the sampled frames.

Performing stratified sampling over streams presents \algname{} with multiple challenges. First, \algname{} must determine how to stratify the stream records. It then must decide how best to allocate its sampling budget across these strata. Finally, in order to compute unbiased estimates for each stratum, \algname{} must draw an unbiased sample from the stream without knowing ahead of time how many records will fall in each stratum. \algname{} overcomes these challenges through the use of sampling techniques which we define and provide intuition for in Section 3.1.

We analyze \algname{} from a theoretical perspective and show that its allocation strategy and expected error converge at quantitative rates. We first derive the per-segment optimal allocation of our sampling budget assuming perfect knowledge of quantities such as the stratum standard deviations and predicate positive rates. We then show that \algname{}'s per-segment sample allocation strategy converges to the optimal allocation at a quantitative rate on stationary streams of data. We further show that the expected error of \algname{}'s estimator converges to zero at a rate inversely proportional to the size of our sampling budget on such streams.

We evaluate \algname{} on six real-world video and text datasets and compare it against two streaming baselines---uniform sampling and stratified sampling with fixed strata and fixed sample allocations. We also compare it against ABae \cite{kang2021abae}, a state-of-the-art algorithm for the batch setting which provides answers to aggregation queries with valid confidence intervals. We show that \algname{} can achieve the same root mean squared error (RMSE) as the streaming baselines with up to 5.0x fewer samples, and it can achieve up to 1.9x lower RMSE than ABae at a fixed number of oracle invocations. We demonstrate these performance improvements on evaluation queries with and without a predicate. We perform a lesion study which shows that each component of \algname{} is critical for it to achieve high performance. We further demonstrate that \algname{}'s  improvement over baselines is not sensitive to the setting of its most significant free parameters. We analyze InQuest's cost and accuracy improvements, as well as the effect that proxy quality has on its evaluation results. Finally, we show that \algname{} is resilient to sudden shifts in the stream parameters: on a set of 100 synthetic datasets that we constructed in an adversarial fashion, \algname{} outperforms our streaming baselines on the RMSE metric by 1.13x-1.42x and performs within 0.99x-1.03x of ABae.

In summary, our paper makes the following contributions:
\begin{enumerate}
    \item We propose an algorithm for optimizing aggregation queries over multiple modalities of unstructured streams of data.
    \item We analyze the algorithm and show that its sample allocation and expected error converge at quantitative rates under certain assumptions.
    \item We evaluate the algorithm on a set of real-world video and text datasets and demonstrate significant improvement over baselines on the RMSE metric.
    \item We demonstrate that even when our theoretical assumptions do not hold, \algname{} empirically outperforms our streaming baselines and is competitive with a state-of-the-art batch setting algorithm.
\end{enumerate}

%% file: tex/overview.tex
\section{Overview and Query Semantics}
We present an overview of the queries that \algname{} optimizes. We first describe our target problem setting and specify our query syntax and semantics. We then provide example queries before defining formal notation for our problem setting.

\subsection{Overview}

\textbf{Target setting.} \algname{} targets streaming aggregation queries with or without a predicate. We assume the query's statistic of interest and predicate (if present) can be computed directly by the oracle or easily derived from its output(s). \algname{} supports streaming queries using the \texttt{AVG}, \texttt{SUM}, or \texttt{COUNT} aggregations. 

\minihead{Proxies}
We assume the user provides a proxy model which returns a real number in some bounded range (e.g., [0, 1]). \algname{} makes no assumptions about proxy quality, but proxies that are more correlated with the target statistic will generate more accurate query results. These proxies can be orders of magnitude cheaper to execute than the oracle (e.g., over 4,000 frames-per-second (fps) for the proxy compared to 3 fps for the oracle \cite{kang2020jointly}). Thus, we make the standard assumption that proxies can be computed in an online fashion over the entire stream without buffering \cite{kang2018blazeit, canel2019scaling}.

\subsection{Query Syntax and Semantics}
\begin{figure}
\centering
\begin{lstlisting}[
  frame=tb,
  basicstyle=\LSTfont,
  keywordstyle=\color{blue},
  language=SQL,
  escapechar=\&
]
SELECT { AVG | SUM | COUNT } ({field | EXPR(field)})
FROM streaming_dataset
[WHERE filter_predicate]
TUMBLE(column, interval) &\Comment{Tumbling window to define segment length}&
ORACLE LIMIT o  &\Comment{oracle invocations per segment}&
[DURATION interval] &\Comment{duration for non-continuous queries}&
USING proxy
\end{lstlisting}
\caption{Syntax for \algname{} which is based on Apache Flink SQL syntax. Users provide a statistic to compute, a dataset, a segment length defined by a tumbling window, an oracle limit per segment, and a proxy model for computing proxy scores in real-time. Users may optionally provide a predicate and/or a query duration (for non-continuous queries).} 
\label{fig:query-syntax}
\end{figure}

\noindent
We show the query syntax for \algname{} in \autoref{fig:query-syntax}. We model our syntax after the Apache Flink SQL language with some minor extensions \cite{katsifodimos2016flink}. Similar to unstructured AQP systems \cite{kang2018blazeit, kang2021abae, kang2017noscope}, a user provides \algname{} with a sampling budget, a proxy model, and an oracle. The user additionally may specify a statistic (i.e., an expression) to compute on each record and an aggregation function (one of \texttt{AVG}, \texttt{SUM}, or \texttt{COUNT}). We assume that any statistic provided by the user is cheap to compute given the output of the oracle. \algname{} also requires the user to specify a tumbling window \cite{flink2023syntax}, whose \texttt{interval} defines the length of each segment, along with a budget of oracle invocations per segment. The \texttt{column} for the tumbling window may be a time-based column or a column specifying each record's index in the stream. The \texttt{interval} can similarly be a time-based range (e.g., \texttt{INTERVAL '1' HOUR}) or it can specify a number of stream records (e.g., \texttt{INTERVAL 10,000 FRAMES}). Finally, \algname{} extends the Apache Flink SQL syntax by allowing users to specify a \texttt{DURATION} for non-continuous queries \cite{babu2001continuous}.



Given these inputs, \algname{} computes an approximate answer to the query. \algname{} aims to provide answers that minimize the mean squared error (MSE) between the approximate result and the ground-truth query result. While higher quality proxies will lead to more accurate query answers, \algname{} will produce an estimate regardless of proxy quality. Query answers can be provided in real-time for both continuous and non-continuous queries, although non-continuous queries will have a final answer provided at the end of the specified \texttt{DURATION}.

\subsection{Examples}
\label{section:examples}
\minihead{Traffic analysis} Consider an urban planner that would like to monitor traffic at an intersection in real-time. The planner wishes to know the per-frame average number of cars that pass through an intersection. The planner could submit the following continuous query to \algname{}:
\begin{lstlisting}[
  frame=tb,
  basicstyle=\LSTfont,
  keywordstyle=\color{blue},
  language=SQL
]
SELECT AVG(count(car)) FROM video
TUMBLE(frame_idx, INTERVAL '108,000' FRAMES)
ORACLE LIMIT 1,000
USING proxy_count_cars(frame)
\end{lstlisting}
where \texttt{count(car)} is computed using an objection detection DNN and \texttt{proxy\_count\_cars} could be computed via an embedding index for unstructured data \cite{kang2020tasti}. In this setting, \texttt{proxy\_count\_cars} returns an estimate of the car count for every frame. The user specifies that each segment should span 108,000 frames (i.e., one hour at 30 fps) and receive a budget of 1,000 oracle invocations.

\minihead{Twitter Sentiment} Consider a journalist that is interested in understanding public sentiment during a presidential debate. For example, the journalist may wish to compute the total number of tweets with positive sentiment that mention a specific candidate. The journalist can submit the following query:

\begin{lstlisting}[
  frame=tb,
  basicstyle=\LSTfont,
  keywordstyle=\color{blue},
  language=SQL
]
SELECT COUNT(positive(tweet)) FROM twitter
TUMBLE(tweet_timestamp, INTERVAL '30' MINUTES)
WHERE mentions_candidate(tweet)
ORACLE LIMIT 5,000
DURATION INTERVAL '4' HOURS
USING proxy_mentions_candidate_pos(tweet)
\end{lstlisting}
The \texttt{mentions\_candidate} predicate is used to filter for tweets that mention the candidate of interest. A large NLP model such as BERT \cite{devlin2018bert} could be used to compute the predicate and the sentiment of the tweet. The proxy could be computed using a smaller NLP model (e.g., fasttext \cite{bojanowski2016enriching}) which would generate a probability in [0, 1] that the tweet mentions the candidate in a positive manner. Since the presidential debate (and post-debate analysis) will only last approximately 4 hours, the user also specifies a \texttt{DURATION}. 



\subsection{Query Formalism}
Formally, let $\mathcal{D} = \{x_i\}$ be a streaming dataset of records and let $O(x_i)$ be the oracle predicate. By definition, $O(x_i) \in \{0, 1\}$ in the predicate case and $O(x_i) = 1$, $\forall x_i \in \mathcal{D}$ in the case without a predicate. We define $\mathcal{D}^{+} = \{x \in \mathcal{D} : O(x) = 1 \}$ to be the subset of the stream that satisfies the query predicate. We further define $X_i = f(x_i) \in \mathbb{R}$ to be the expression the query aggregates over, $N$ to be the per-segment sampling budget, and $T$ to be the number of processed segments.

\algname{} computes $\mu = \sum_{x \in \mathcal{D}^{+}} f(x)/|\mathcal{D}^{+}$| via an approximation $\hat{\mu}$, with its total sampling budget $N T$ up to the current segment. We measure query result quality by the MSE, i.e., $|\mu - \hat{\mu}|^2$.

%% file: tex/algorithm.tex
\section{\algname{} Description and Query Processing}
We describe \algname{} for accelerating aggregation queries on streams of unstructured data. 
We first provide intuition for \algname{}'s design and define relevant sampling terminology. We then discuss the challenges of the problem setting before providing an overview of how \algname{} addresses these challenges. Finally, we provide the pseudocode for \algname{} and its subroutines. Formal notation used throughout this section is presented in \autoref{tab:notation}.

\subsection{Background and Algorithm Intuition}
\begin{figure}[t!]
  \centering
  \includegraphics[width=\linewidth]{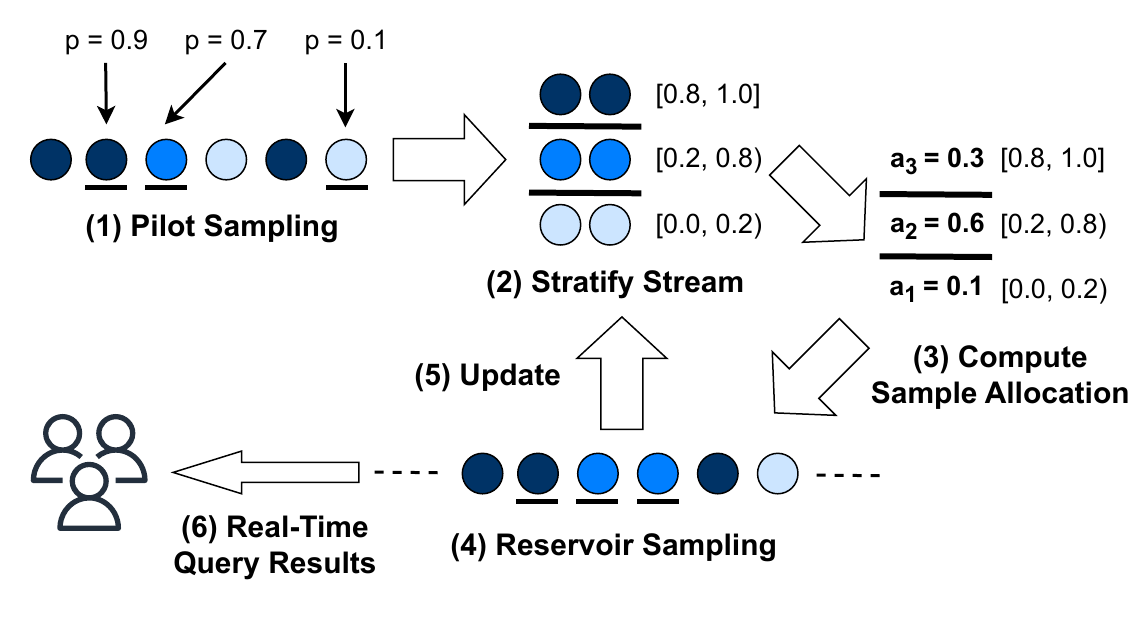}
  \caption{\algname{}'s high-level workflow. By design, \algname{} uses stratified sampling to reduce the variance of its query estimates. It uses pilot sampling to compute an initial stratification and then updates it using the history of oracle samples. Users can extract query estimates in real-time.}
  \label{fig:inquest-workflow-diagram}
\end{figure}

Our first design decision for \algname{} was to leverage \textit{stratified sampling} \cite{parsons2014stratified} to compute precise query estimates with standard frequentist bounds. Stratified sampling is a method in which the target population (i.e., the stream) is divided into distinct sub-populations (i.e., strata) which are sampled from independently. The objective is to stratify the target population such that elements in each stratum are similar to one another in terms of a statistic of interest. An estimate over the entire population can then be computed with smaller error (relative to uniform sampling) by aggregating lower variance estimates from each stratum.

In order to perform stratified sampling over streams, we must first determine the boundaries of our strata (in terms of proxy scores). As illustrated in the first step of \autoref{fig:inquest-workflow-diagram}, \algname{} performs \textit{pilot sampling} \cite{browne1995use} to accomplish this task. Pilot sampling is a technique in which a fraction of one's sampling budget is used to produce an initial estimate of some quantity (in our case, the ideal strata boundaries).

Once \algname{} has constructed initial strata (\autoref{fig:inquest-workflow-diagram}, step 2) it needs to allocate its sampling budget efficiently across the strata. There is a known optimal allocation for stratified sampling (see Section 4.2), but computing it requires perfect knowledge of quantities such as the strata standard deviations. In light of this, \algname{} approximates the optimal allocation using the oracle predictions from its predicate matching pilot samples (\autoref{fig:inquest-workflow-diagram}, step 3).

Estimating the optimal allocation using pilot samples comes with risk, because variance in the pilot sample may result in sampling few (or potentially 0) predicate matching samples in one or more strata. This outcome is considerably more likely in the stream setting, where the number of predicate matching samples can fluctuate as a function of time. If no predicate matching samples are drawn in a stratum, the optimal allocation would assign 0 samples to that stratum, thus leading to catastrophic under-allocation. \algname{} uses \textit{defensive sampling} \cite{owen2000safe} to protect against this outcome. In this context, defensive sampling is the practice of allocating a fraction of one's sampling budget evenly across all strata to ensure a minimum number of samples is allocated to each stratum (\autoref{fig:inquest-workflow-diagram}, step 3).

Once its per-stratum sampling budget is allocated, \algname{} must finally determine which samples to draw from the stream. This is non-trivial in the stream setting, because \algname{} cannot know ahead of time which records will fall in each stratum. A naive solution to this problem would be to greedily sample records that fall in each stratum until the sampling budget is exhausted. However, this would produce a sample that is biased towards the beginning of the stream. To overcome this issue, \algname{} makes use of \textit{reservoir sampling} \cite{aggarwal2006biased, al2007adaptive}. Reservoir sampling is a technique that is guaranteed to sample stream records uniformly in time, without prior knowledge of the stream length. This enables \algname{} to produce unbiased samples for each stratum in the stream (\autoref{fig:inquest-workflow-diagram}, step 4).

Stream parameters, such as the strata standard deviations, can shift over time. Thus, \algname{} processes the stream in segments and updates its stratification and sample allocation at the end of each segment (\autoref{fig:inquest-workflow-diagram}, step 5). Finally, a user may retrieve \algname{}'s latest query estimate at any point in time (\autoref{fig:inquest-workflow-diagram}, step 6).

\begin{table}[t!]
    \caption{Summary of notation.}
    \label{tab:notation}
    \begin{tabular}{ll}
        Symbol & Description \\
        \hline
        $\mathcal{D}$ & Streaming dataset of records \\
        $\mathcal{S}$ & Stratification, i.e., $k$ strata \\
        $\mathcal{P}(x)$ & Proxy model \\
        $T$ & Number of segments (including pilot segment) \\
        $N$ & Per-segment user-specified sampling budget \\
        $N_{1}$ & Per-segment defensive sample budget \\
        $N_{2}$ & Per-segment dynamic sample budget \\
        $K$ & Number of strata \\
        $\mathcal{O}(x)$ & Oracle predicate \\
        $\mathcal{D}_{tk}$ & Set of dataset records in segment $T$ and stratum $k$\\
        $X_{tk,i}$ & $i$th sample from $\mathcal{D}_{tk}$ \\
        $X_{tk}$ & Set of samples drawn from $\mathcal{D}_{tk}$ \\
        $X_{tk}^{+}$ & Set of predicate matching samples drawn from $\mathcal{D}_{tk}$ \\
        $p_{tk}$ & Predicate positive rate \\
        $w_{tk}$ & $|D_{tk}| p_{tk}$/$\sum |D_{tj}| p_{tj}$ \\
        $\sigma_{tk}$ & True std. dev. of the samples in $\mathcal{D}_{tk}$ \\
        $a^{*}_{tk}$ & Optimal fraction of $N_2$ allocated to $\mathcal{D}_{tk}$ \\
        $f(x)$ & Statistic function
    \end{tabular}
\end{table}

\subsection{\algname{} Algorithm}
\minihead{Challenges} Our problem setting involves a number of key challenges. Similar to some prior work \cite{kang2021abae}, we do not know the correlation between the proxy model $\mathcal{P}(x)$ and the ground-truth statistic function $f(x)$ ahead of time. We also do not have prior knowledge of the quantities $\sigma_{tk}$ and $p_{tk}$. This prevents us from using standard AQP techniques \cite{piatestsky-shapiro1984accurate,braverman2014generalizing} to leverage this information and pre-compute an optimal allocation of our sampling budget. 

The streaming nature of our problem creates additional challenges. In particular, the distributions of $\mathcal{P}(x)$ and $f(x)$, and the related quantities $\sigma_{tk}$ and $p_{tk}$, are also a function of time. Thus, even if we calibrate our sample allocation based on the history of these distributions, we have no guarantees that these distributions will not change in the future.

\begin{algorithm}[t!]
\caption{Pseudocode for \algname{}. \algname{} performs pilot sampling to initially stratify the dataset. It then performs stratified reservoir sampling on each segment as it iteratively updates $\hat{a}_{tk}$}
\label{alg:algo}
\begin{algorithmic}[1]

\Function{\algname{}Pilot}{$\mathcal{D}_1$, $N_{\textrm{pilot}}$, $K$}
\State $X_1 \gets \text{UniformSampling}(D_1, N_{\textrm{pilot}})$
\State $X_1^{+} \gets \{x \vert x \in X_1, \mathcal{O}(x) = 1 \}$
\State \Return $X_1$, $X_1^{+}$
\EndFunction \\

\Function{\algname{}}{$\mathcal{D}$, $\mathcal{O}$, $\mathcal{P}$, $K$, $N_1$, $N_2$}
\State $X_1, X_1^{+} \gets \Call{\algname{}Pilot}{\mathcal{D}_1, N_{1} + N_{2}, K}$
\For{$t \in [2, 3, \dots]$}
    \State $\mathcal{\hat{S}}_{t} \gets$ \Call{GetStrata}{$\mathcal{P}$, $\mathcal{D}_{t-1}$, $K$, $\alpha$, $\mathcal{S}_{<t}$}
    \State $\hat{a}_{t} \gets$ \Call{GetAlloc}{$\mathcal{D}_{t-1}$, $K$, $N_1$, $N_2$, $a_{<t}$, $X_{t-1}$, $X_{t-1}^{+}$}
    \State $\mathcal{D}_{t1}, \dots, \mathcal{D}_{tK} \gets \textrm{SplitStream}(\mathcal{D}_t, \mathcal{\hat{S}}_t)$
    \For{$k \in [1, \dots, K]$}
        \State $X_t \gets \textrm{ReservoirSampling}(\mathcal{D}_{tk}, N \hat{a}_{tk})$
        \State $X_t^{+} \gets \{x \vert x \in X_t, \mathcal{O}(x) = 1 \}$
    \EndFor
\EndFor
\State $\hat{\mu} \gets$ \Call{GetPrediction}{$X_{1}$, $\dots$, $X_{T}$, $X_{1}^{+}$, $\dots$, $X_{T}^{+}$, $\mathcal{D}$}
\State \Return $\hat{\mu}$
\EndFunction
\end{algorithmic}
\end{algorithm}

\begin{algorithm}[t!]
\caption{Subroutines for \algname{}.}
\label{alg:algo2}
\begin{algorithmic}[1]

\Function{GetStrata}{$\mathcal{P}$, $\mathcal{D}_{t-1}$, $K$, $\alpha$, $\mathcal{S}_{<t}$}
\State $\mathcal{S}_{t-1} \gets \textrm{StratifyByQuantile}(\mathcal{P}(\mathcal{D}_{t-1}), K)$
\State $\mathcal{\hat{S}}_{t} \gets \textrm{EWMA}(\{\mathcal{S}_1, \dots, \mathcal{S}_{t-1}\}, \alpha)$
\State \Return $\mathcal{\hat{S}}_{t}$
\EndFunction \\

\Function{GetAlloc}{$\mathcal{D}_{t-1}$, $K$, $N_1$, $N_2$, $a_{<t}$, $X_{t-1}$, $X_{t-1}^{+}$}
\For{$k \in [1, \dots, K]$}
    \State $\hat{p}_{t-1,k} \gets \frac{|X_{t-1,k}^{+}|}{|X_{t-1,k}|}$
    \State $\hat{\mu}_{t-1,k} \gets \frac{\sum_{x \in X_{t-1,k}^{+}} f(x)}{|X_{t-1,k}^{+}|} \textrm{ if } |X_{t-1,k}^{+}| > 0 \textrm{ else } 0$
    \State $\hat{\sigma}_{t-1,k}^2 \gets \frac{\sum_{x \in X_{t-1,k}^{+}} (f(x) - \mu_{t-1,k})^2}{(|X_{t-1,k}^{+}| - 1)} \textrm{ if } |X_{t-1,k}^{+}| > 1 \textrm{ else } 0$
    \State $\hat{w}_{t-1,k} \gets \sqrt{\hat{p}_{t-1,k}} \cdot \frac{|\mathcal{D}_{t-1,k}|}{|\mathcal{D}_{t-1}|}$
\EndFor
\For{$k \in [1, \dots, K]$}
    \State $a_{t-1,k} \gets \frac{\hat{w}_{t-1,k} \hat{\sigma}_{t-1,k}}{\sum_{j=1}^K \hat{w}_{t-1,j} \hat{\sigma}_{t-1,j}}$
\EndFor
\State $\hat{a}_{t} \gets \textrm{EWMA}(\{a_{1}, \dots, a_{t-1}\}, \alpha)$
\For{$k \in [1, \dots, K]$}
    \State $\hat{a}_{tk} \gets \frac{N_1/K + N_2 \hat{a}_{tk}}{N}$
\EndFor
\State \Return $\hat{a}_{t}$
\EndFunction \\

\Function{GetPrediction}{$X_{1}$, $\dots$, $X_{T}$, $X_{1}^{+}$, $\dots$, $X_{T}^{+}$, $\mathcal{D}$}
\For{$t \in [1, T]$}
    \For{$k \in [1, K]$}
        \State $\hat{p}_{tk} \gets \frac{|X_{tk}^{+}|}{|X_{tk}|}$
        \State $\hat{\mu}_{tk} \gets \frac{\sum_{x \in X_{tk}^{+}} f(x)}{|X_{tk}^{+}|}$
    \EndFor
\EndFor
\State $\hat{\mu} \gets \sum_{t=1}^T \sum_{k=1}^K \hat{\mu}_{tk} \cdot \frac{\hat{p}_{tk}|\mathcal{D}_{tk}|}{\sum_{t=1}^T \sum_{j=1}^K \hat{p}_{tj}|\mathcal{D}_{tj}|}$
\State \Return $\hat{\mu}$
\EndFunction
\end{algorithmic}
\end{algorithm}

\minihead{Overview} The challenges highlighted above present unique difficulties for \algname{}. The optimal allocation of our sampling budget $N$ to the strata $\mathcal{D}_{tk}$ depends on the per-strata standard deviations of the ground-truth statistic and predicate positivity rates \cite{optimalallocation}. Without prior knowledge of $\sigma_{tk}$ and $p_{tk}$ we cannot directly compute the optimal stratified sampling allocation $a^{*}_{tk}$. Instead, we must estimate $\sigma_{tk}$, $p_{tk}$, and $a^{*}_{tk}$ using previously drawn samples. Furthermore, the standard deviations $\sigma_{tk}$ and predicate positive rates $p_{tk}$ can vary from segment-to-segment. This means that our estimates of $\sigma_{tk}$, $p_{tk}$, and $a^{*}_{tk}$ are susceptible to distribution shifts in the stream of data. Finally, in standard stratified sampling it is often beneficial to stratify the dataset such that each stratum contains a roughly equal number of records. Since we do not know the distribution of proxy values ahead of time, it is difficult (without buffering the entire stream) for us to construct our stratification $\mathcal{S}_{tk}$ such that each stratum will contain an equal number of records.

To address this, \algname{} trades off learning from the history of the stream with the need to adapt to shifts in the distribution of proxy values, $\sigma_{tk}$, and $p_{tk}$ across segments. \algname{} does this by updating its stratification and sample allocation under a weighted moving average. Additionally, \algname{} reserves defensive samples to increase its resilience to extreme shifts in the distributions of $\mathcal{P}(x)$ and $f(x)$. On some datasets we found that \algname{} could suffer catastrophic failures without defensive sampling. Specifically, if the sample standard deviation $\hat{\sigma}_{tk}$ was close to 0 (or if no predicate matching records were sampled in the predicate setting), \algname{} could undersample a stratum for the remainder of the query.


\minihead{Formal description}
Recall our notation from \autoref{tab:notation}. In particular, note that $\mathcal{O}(x)$ is the oracle predicate, $P(x)$ is the proxy predicate, and $\mathcal{D}$ is our streaming dataset of records. We define $\mathcal{D}_{tk}$ to be the subset of records in segment $t$ and stratum $k$. Finally, we denote $X_{tk,i}$ and $X_{tk,i}^{+}$ to be the $i$th sample and the $i$th predicate matching sample drawn from $\mathcal{D}_{tk}$, respectively.

The free parameters of \algname{} include its sampling budgets ($N_1$ and $N_2$), the number of strata ($K$), and the smoothing parameter for its weighted moving averages ($\alpha$). \algname{} will also compute several other quantities, including sample means, predicate positive rates, and allocations ($\hat{\mu}_{tk}$, $\hat{p}_{tk}$, and $\hat{a}_{tk}$).


\algname{} uniformly samples $N$ samples from the pilot fraction of the query. It then processes each query segment by first updating its stratification and sample allocation before performing reservoir sampling. We present the pseudocode for \algname{} in the predicate setting in Algorithm \autoref{alg:algo}. The pseudocode for the no predicate setting can be recovered by setting $X_{tk} = X_{tk}^{+}$ and $p_{tk} = 1$ for all $t$ and $k$.


\algname{} computes its stratification $\mathcal{\hat{S}}_t$ in the \texttt{GetStrata} subroutine in Algorithm 2. \algname{} first stratifies the previous segment's samples by proxy value quantile, such that $1/K$ records in the previous segment fall in each strata. It then updates $\mathcal{\hat{S}}_t$ to be the exponential weighted moving average of the history of $\mathcal{S}_1$, $\dots$, $\mathcal{S}_{t-1}$. The aggressiveness of the weighted moving average is controlled by the smoothing parameter $\alpha$.

Next, \algname{} computes its sample allocation $\hat{a}_{t}$ in the \texttt{GetAlloc} subroutine in Algorithm 2. \algname{} first computes the previous segment's sample standard deviations $\hat{\sigma}_{t-1,k}$ and predicate positive rates $\hat{p}_{t-1,k}$. It then computes the optimal allocation $a_{t-1}$ as a weighted average of the stratum standard deviations. \algname{} then computes the sample allocation $\hat{a}_{t}$ to be the exponential weighted moving average of the history of $a_1$, $\dots$, $a_{t-1}$. Finally, it adjusts $\hat{a}_t$ to include $N_1/K$ defensive samples per stratum.

Once the stratification and sample allocation are computed, \algname{} performs reservoir sampling in each $\mathcal{D}_{tk}$. \algname{} repeats this process for each segment before finally computing its prediction $\hat{\mu}$ as a weighted average of the sample means $\hat{\mu}_{tk}$ in the \texttt{GetPrediction} subroutine in Algorithm 2.

\minihead{Setting parameters} By default, \algname{} uses the parameter settings: $K = 3$, $\alpha = 0.8$, and $N_1 = 0.1$. In Section 5 we demonstrate that these parameters achieve strong performance results on six real-world and 100 synthetic datasets, relative to both streaming and batch setting algorithms. Advanced users may optionally tune these parameters to optimize performance on their own datasets. As a general guideline, we recommend setting $N_1$ to be small ($\sim$ 5-10\% of $N$) and setting $N$ and $K$ such that one would reasonably expect to get at least 20-100 samples per segment and stratum.


\minihead{Confidence interval} We use the bootstrap to compute CIs. One method for showing that the bootstrap is valid is to demonstrate its asymptotic validity. The asymptotic validity of sampling with stochastic draws follows from the analysis in \cite{abaetechreport}. We can also use a standard subgaussian tail bound, but they give similar results.

%% file: tex/analysis.tex
\section{Theoretical Analysis}
We analyze \algname{} from a theoretical perspective and show that its allocation and expected error converge at quantitative rates for stationary streams. We first show that \algname{}'s sample allocation converges to the optimal stratified sampling allocation at a rate $O\Big(\frac{1}{N_1(t-1)}\Big)$. We then show that \algname{}'s expected MSE converges to zero at a rate $O\Big(\frac{1}{N_1} + \frac{N_1}{N_2^2} + \frac{1}{N_2\sqrt{N_1}} + \frac{1}{N_2\sqrt{N_1 t}} \Big)$.

\subsection{Notation and Preliminaries}
\label{section:assumptions}
\minihead{Notation} Recall our notation in \autoref{tab:notation}. We denote $p_{tk}$ and $\sigma_{tk}$ to be the predicate positive rate and standard deviation of $\mathcal{D}_{tk}$ respectively. We further define $\mu_{t}$ to be the segment mean, $w_{tk}$ to be the fraction of dataset records that fall in $\mathcal{D}_{tk}$, and $a_{tk}$ to be the allocation of our sampling budget $N$.

\minihead{Assumptions} For subsections 4.3 and 4.4, we assume that our streaming dataset follows a stationary distribution. Specifically, we assume that:
\begin{align}
    \sigma_{tk} &= \sigma_{rk} : \forall t, r \in T \\
    p_{tk} &= p_{rk} : \forall t, r \in T \\
    w_{tk} &= w_{rk} : \forall t, r \in T
\end{align}

While these assumptions may not hold true for real-world datasets, they are important for making our proofs tractable. In Sections 5.2 and 5.6 we present empirical evidence that \algname{} performs well relative to baselines even when these assumptions break down.

Recall that $X_{tk,i}$ is the $i$th predicate matching sample drawn from $\mathcal{D}_{tk}$. We assume that $X_{tk,i}$ is a sub-Gaussian random variable with nonzero standard deviation. This enables us to upper bound functions that sum sub-Gaussian variables (e.g., $\mu_{tk}$ and $\sigma_{tk}^2$) with constants such as $C^{\mu_{tk}}$ and $C^{\sigma_{tk}^2}$. We further assume that at least one stratum has non-zero $p_{tk}$.

\subsection{Optimal Stratified Sampling Allocation with Perfect Information}
We begin by analyzing the optimal allocation of our dynamic sample budget $N_2$ in segment $t$. We assume perfect knowledge of $\sigma_{tk}$ and $p_{tk}$ and that we deterministically draw $|X_{tk}^{+}| = p_{tk} \big(\frac{N_1}{K} + N_2 a_{tk} \big)$ samples from each stratum. We present the analysis for the setting with a predicate, but note that these results also hold for the no predicate setting where $p_{tk} = p_t = 1$.

\begin{proposition}
Assume that $\sigma_{tk}$ is known and we draw $|X_{tk}^{+}| = p_{tk}\big(\frac{N_1}{K} + N_2 a_{tk}\big)$ samples per stratum in segment $t > 2$ (up to rounding effects). Then the choice $a_{tk} = a_{tk}^{*}$ that minimizes the MSE of the unbiased estimator $\hat{\mu}_{t} = \sum_{k=1}^K w_{tk} \cdot \frac{\sum_{ x \in X_{tk}^{+}} f(x)}{|X_{tk}^{+}|}$ is:

\begin{equation}
    a_{tk}^{*} = \frac{|\mathcal{D}_{tk}| \sqrt{p_{tk}} \sigma_{tk}}{\frac{N_2}{N} \sum_{j=1}^K |\mathcal{D}_{tj}| \sqrt{p_{tj}} \sigma_{tj}} - \frac{N_1}{N_2 K}
\end{equation}
\end{proposition}

\begin{proposition}
Suppose the conditions in Proposition 1 hold. Then the expected MSE of the estimator $\hat{\mu_t}$ under the allocation $a_{tk}^{*}$ is

\begin{align}
    \mathbb{E}[(\hat{\mu}_{t}^{*} - \mu_{t})^2] &= \sum_{k=1}^K \frac{w_{tk}^2 \sigma_{tk}^2}{p_{tk} \big(\frac{N_1}{K} + N_2 a_{tk}^{*} \big)} \\
    &= \frac{1}{N p_{all}^2} \sum_{k=1}^K |\mathcal{D}_{tk}| \sqrt{p_{tk}} \sigma_{tk} \Bigg(\sum_{j=1}^K |\mathcal{D}_{tj}| \sqrt{p_{tj}} \sigma_{tj} \Bigg)
\end{align}
Where $p_{all}$ is defined as
\begin{equation}
    p_{all} = \sum_{j=1}^K |\mathcal{D}_{tj}| p_{tj}
\end{equation}
\end{proposition}

Our expression for $a_{tk}^{*}$ shows that the optimal allocation is weighted towards strata with larger $|\mathcal{D}_{tk}|$, $p_{tk}$, and $\sigma_{tk}$. Intuitively, we want to spend our sampling budget on strata that are more likely to contain predicate matching records. Furthermore, strata with greater $\sigma_{tk}$ will generally require more samples to get an accurate estimate of $\mu_{tk}$.

The expression for the expected error shows that larger strata standard deviations will lead to an increase in the error. The expected error also increases inversely with respect to $p_{all}^2$, where $p_{all}$ is a weighted sum of the strata predicate positive rates. Intuitively, as $p_{all}$ goes to 0 it becomes harder for \algname{} to compute accurate estimates of $\mu_{tk}$ because it becomes increasingly unlikely that \algname{} will draw a sample in $X_{tk}$ that matches the predicate. Finally, the expected error decreases linearly with respect to our sampling budget $N$.

\subsection{\algname{} Sample Allocation Converges to Optimal Stratified Sampling Allocation}
We analyze \algname{}'s dynamic sample allocation and prove that it converges to the optimal allocation at a quantitative rate under the assumptions stated in \autoref{section:assumptions}. For this analysis we further assume that $\alpha = 0$, i.e., we compute the update to the sample allocation $\hat{a}_{tk}$ based on the unweighted history of the samples. We provide the theorem statement but defer the full proof to an extended technical report \cite{techreport}.

\begin{theorem}
Under the assumptions stated in \autoref{section:assumptions} and with high probability over the samples drawn in segments $[2, \dots, t]$
\begin{equation}
\label{eq:alloc-rate}
    \mathbb{E}[(\hat{a}_{tk} - a_{tk}^{*})^2] \leq O\Big(\frac{1}{N_1(t-1)}\Big)
\end{equation}
\end{theorem}

\noindent
\autoref{eq:alloc-rate} shows that \algname{}'s sample allocation converges to the optimal allocation at a rate that decreases linearly as a function of $N_1$ and $t$. The product $N_1(t-1)$ represents the total number of defensive samples in all segments leading up to segment $t$.

\subsection{\algname{} Error Convergence}
We analyze \algname{}'s expected error and prove that it converges at a quantitative rate under the assumptions stated in \autoref{section:assumptions}. For this analysis we further assume that $\alpha = 0$. We provide the theorem statement, but once again defer the full proof to an extended technical report \cite{techreport}.

\begin{theorem}
Under the assumptions stated in \autoref{section:assumptions} and with high probability over the samples drawn in segments $[1, \dots, t]$
\begin{equation}
    \mathbb{E}[(\hat{\mu}_t - \mu_t)^2] \leq O\Big(\frac{1}{N_1} + \frac{N_1}{N_2^2} + \frac{1}{N_2\sqrt{N_1}} + \frac{1}{N_2\sqrt{N_1 t}}\Big)
\end{equation}
Furthermore, if $N_1 = N_2$ then this simplifies to
\begin{equation}
    \mathbb{E}[(\hat{\mu}_t - \mu_t)^2] \leq O\Big(\frac{1}{N}\Big)
\end{equation}
\end{theorem}

\subsection{Understanding \algname{}}
We provide proof sketches for the theorems and discuss some aspects of the analysis of \algname{} that are of broader interest. \\


\noindent
\textit{4.3.1 Proof Sketch: \algname{} Allocation Convergence.} We use concentration inequalities to bound our random variables, specifically $p_{<tk}$ and $\sigma_{<tk}$. We then compute an upper bound on the expected mean squared error of the difference between $\hat{a}_{tk}$ and $a_{tk}^{*}$. We separately compute the upper bound for cases where $p_{<tk}$ is small (i.e., less than $\frac{1}{N_1}$) and cases where $p_{<tk}$ is large. We simplify the expectation and conclude that our allocation error converges at a rate of $O\Big(\frac{1}{N_1(t-1)}\Big)$. \\

\noindent
\textit{4.4.1 Proof Sketch: \algname{} Error Convergence.} We use concentration inequalities to bound our random variables, including $p_{tk}$ and $\sigma_{tk}$, as well as on other quantities derived from these variables. We use these bounds to compute a high probability lower bound on the number of predicate matching samples drawn for each stratum (i.e., $|X_{tk}^{+}|$) for the case where $p_{tk}$ is large (i.e., larger than $\frac{1}{N_1}$). We then derive the upper bound on the expected error for strata where $p_{tk}$ is large and show that the error for the remaining strata becomes negligible. Finally, we simplify our expectation and conclude that our error converges at a rate of $O\Big(\frac{1}{N}\Big)$. \\



\noindent
\textit{4.4.2 Challenges.} We discuss several challenges in the analysis of \algname{}. Recent work has analyzed using stochastic draws in the batch setting, where the dataset can be stratified perfectly and pilot sampling can be performed with samples drawn from the entire dataset \cite{kang2021abae}. We extend this work using stochastic draws to the stream setting and show that \algname{} can achieve optimal performance on stationary datasets.

\minihead{Estimating key quantities}
Prior work in stratified sampling assumes that features of the data distribution in each stratum, such as $p_{tk}$ and $\sigma_{tk}$, are known \cite{optimalallocation}. It then uses this knowledge to construct optimal sample allocations. In contrast, \algname{} has no prior knowledge of these quantities and must estimate them from samples it draws stochastically. For values of $p_{tk}$ that are small relative to our sample budget $N$ (e.g., $p_{tk} < \frac{1}{N_2}$), \algname{} may not draw a single predicate matching sample, thus making it impossible to accurately estimate $p_{tk}$ and $\sigma_{tk}$. 

\minihead{Stochastic sample sizes}
In the predicate setting \algname{} may sample records that do not satisfy the predicate. As a result, the number of predicate matching samples in each $\mathcal{D}_{tk}$ is stochastic. This is in contrast to standard stratified sampling, which assumes a deterministic number of draws from each stratum. In the case where both $p_{tk}$ and $|X_{tk}|$ are large, \algname{} will draw approximately $p_{tk}|X_{tk}|$ samples which will result in estimates with similar quality to an estimator with $p_{tk}|X_{tk}|$ deterministic samples. However, for small $p_{tk}$ this no longer holds true. 

\minihead{Recursive Definitions} By design, \algname{} samples the stream and updates its allocation at discrete intervals throughout the query. The allocation $\hat{a}_{tk}$ depends on the history of samples drawn in segments $[1, \dots, t-1]$. Specifically, it's a function of $\hat{p}_{tk}$ and $\hat{\sigma}_{tk}$. In turn, $\hat{p}_{tk}$ and $\hat{\sigma}_{tk}$ depend on the number of samples drawn from $\mathcal{D}_{tk}$. This means they are a function of the allocation $\hat{a}_{tk}$. The recursive nature of these definitions makes it challenging to apply meaningful concentration inequalities on these random variables.

%% file: tex/evaluation.tex
\section{Evaluation}
We evaluated our algorithm on six real-world video and text datasets. We first describe our experimental setup and baselines. We then show that \algname{} outperforms the stream setting baselines on all datasets we consider, achieving the same root mean squared error (RMSE) with up to 5.0x fewer samples. We further demonstrate that \algname{} outperforms ABae \cite{kang2021abae}---a state-of-the-art algorithm for the batch setting---by up to 1.9x on the RMSE metric at a fixed sample budget. We then show that each of \algname{}'s major components contributes to its performance and that it is not sensitive to the setting of its parameters. Finally, we analyze \algname{}'s cost and accuracy improvements, examine effect that proxy quality has on its evaluation results, and demonstrate that \algname{} is resilient to rapid changes in the stream parameters.

\subsection{Experimental Setup}
\textbf{Datasets, proxies, and oracles.} We considered six real-world video and text datasets (\autoref{tab:datasets}). The video datasets are commonly used for video analytics evaluation \cite{kang2018blazeit, kang2021abae, kang2017noscope, koudas2020video}. The text dataset is publicly available on Kaggle and contains 3M+ tweets between users and customer support Twitter accounts \cite{kaggle2022dataset}. For each video dataset we generated proxy scores from TASTI embeddings that we created with a pre-trained ResNet-18 model \cite{kang2020tasti, he2015resnet}. Our oracle labels were computed using a Mask R-CNN model \cite{he2017maskrcnn}. For our text dataset we generated proxy scores using a fasttext model, while our oracle labels were computed using a HuggingFace BERT model trained on English language tweets \cite{loureiro2022timelms, huggingface2022model}.

\minihead{Evaluation queries}
We evaluated the baselines and \algname{} on each dataset using two queries, one with a predicate and one without. The queries with a predicate were all of the form:

\begin{lstlisting}[
  frame=tb,
  basicstyle=\LSTfont,
  keywordstyle=\color{blue},
  language=SQL
]
SELECT AVG(expr(record)) FROM dataset
WHERE filter_predicate
TUMBLE(record_idx, INTERVAL '100,000' RECORDS)
ORACLE LIMIT N
DURATION INTERVAL '500,000' RECORDS
USING proxy
\end{lstlisting}

\noindent
For our video datasets, \texttt{expr} was \texttt{count\_boats} for the \texttt{grand-canal} and \texttt{rialto} datasets and \texttt{count\_cars} for the rest. The predicate was \texttt{count\_boats(record) > 0} and \texttt{count\_cars(record) > 0}, respectively. For our text dataset \texttt{expr} was \texttt{sentiment} and the predicate was \texttt{is\_customer\_tweet(record)}. The queries without a predicate were identical to the one shown above with the \texttt{WHERE} clause removed. Each query allocated 10\% of the oracle budget for defensive sampling.


\begin{table}[t!]
  \caption{Summary of datasets, predicates, predicate positivity rates $p$, and proxy correlation to the groundtruth statistic $r$ (Pearson product-moment correlation coefficient).}
  \label{tab:datasets}
  \begin{tabular}{llcc}
    \toprule
    Dataset & Predicate & $p$ & $r$ \\
    \midrule
    \texttt{archie} & At least one car & 0.50 & 0.92 \\
    \texttt{customer-support} & Is customer tweet & 0.56 & 0.79 \\
    \texttt{grand-canal} & At least one boat & 0.60 & 0.91 \\
    \texttt{night-street} & At least one car & 0.37 & 0.92 \\
    \texttt{rialto} & At least one boat & 0.89 & 0.91 \\
    \texttt{taipei} & At least one car & 0.63 & 0.87 \\
    \bottomrule
  \end{tabular}
\end{table}

\minihead{Streaming methods evaluated}
We compared our algorithm against two baselines for the stream setting: uniform sampling and stratified sampling with fixed strata and fixed sample allocations. For uniform sampling, we precomputed a set of $N$ frames (where $N$ is the oracle budget) to sample between the query submission time and the end of the query's \texttt{DURATION}. We then called the oracle on these records and computed our query estimate by averaging the per-record statistic on the sampled frames:

\begin{lstlisting}[
  frame=tb,
  basicstyle=\LSTfont,
  keywordstyle=\color{blue},
  language=SQL
]
SELECT AVG(expr(record)) FROM uniform_sample(dataset)
ORACLE LIMIT N
DURATION INTERVAL '500,000' RECORDS
\end{lstlisting}

\noindent
For queries with a predicate, the estimate was only computed using the statistic values from the predicate-matching samples:

\begin{lstlisting}[
  frame=tb,
  basicstyle=\LSTfont,
  keywordstyle=\color{blue},
  language=SQL
]
SELECT AVG(expr(record)) FROM uniform_sample(dataset)
WHERE expr(record) > 0
ORACLE LIMIT N
DURATION INTERVAL '500,000' RECORDS
\end{lstlisting}

\noindent
For the stratified sampling baseline, we executed evaluation queries similar to the ones used for \algname{}. We also performed stratified sampling within each segment. However, each segment and stratum pair (i.e., $\mathcal{D}_{tk}$) received a fixed oracle budget of $\frac{N}{K}$ samples and maintained a fixed stratification of $k_1 = [0,0.33]$, $k_2 = [0.33,0.67]$, and $k_3 =[0.67, 1.0]$. Due to the streaming nature of our queries, we performed reservoir sampling within each $\mathcal{D}_{tk}$ to ensure that the oracle was applied uniformly at random. We then computed our estimate for each segment as a weighted average of the aggregation function \texttt{AGG} (one of \texttt{AVG}, \texttt{SUM}, or \texttt{COUNT}) applied to the samples from each $\mathcal{D}_{tk}$:

\[ \hat{\mu}_t = \sum_{k=1}^K \hat{w}_{tk} \cdot \textrm{AGG}\big(\{f(x) | x \in S_{tk}\}\big) \]

\noindent
Where the weight $\hat{w}_{tk}$ is the estimate of the fraction of predicate matching samples that fall in $\mathcal{D}_{tk}$:

\begin{align}
    \hat{p}_{tk} &= \frac{|\{x | O(x) = 1, x \in S_{tk}\}|}{|S_{tk}|} \\
    \hat{w}_{tk} &= \frac{|\mathcal{D}_{tk}| \cdot \hat{p}_{tk}}{\sum_{i=1}^K |\mathcal{D}_{ti}| \cdot \hat{p}_{ti}}
\end{align}

\noindent
For queries without a predicate $\hat{p}_{tk} = p_{tk} = 1$ which meant our estimate $\hat{w}_{tk}$ was computed exactly. In our baseline experiments for stratified sampling with fixed strata we set $K=3$ and configured our segment length such that there were $T=5$ segments.

\minihead{Batch methods evaluated} In order to compare \algname{} to prior work, we also evaluated ABae \cite{kang2021abae} using near-identical evaluation queries (minor syntax tweaks were necessary for the batch setting). We chose ABae for the comparison because it also provides approximate answers to aggregation queries with valid confidence intervals. To run the evaluation, we presented our streaming datasets to ABae as if they were batch datasets. We ran ABae with sample reuse, $K=3$, and allocated 15\% of its budget to pilot sampling.

ABae has the advantage of observing the proxy score distribution over the entire dataset, which allows it to compute an optimal stratification and sample allocation before it begins sampling. In contrast, \algname{} does not have the benefit knowing the proxy score distribution prior to sampling. In spite of this, we find the comparison useful for contextualizing \algname{}'s results, and we show that \algname{} outperforms ABae on our key metric of interest.

\minihead{Metrics} Our primary metric of interest is the RMSE between each method's estimate of the expression in the \texttt{SELECT} clause and the oracle value. In particular, we measure the RMSE on each segment of the query and evaluate each method by computing the median RMSE across all query segments. We evaluated the RMSE at different oracle budgets representing 0.1 - 1\% of the total records in each query. We additionally compared the number of samples needed to achieve a fixed error target.

\minihead{Implementation} We implemented our algorithm, baselines, and experimental evaluation in Python. Our open-sourced code can be found at https://github.com/stanford-futuredata/InQuest.

\subsection{\algname{} End-to-end Performance}
We first investigated whether or not \algname{} outperforms our baselines on the median segment RMSE metric. For each dataset we evaluated the uniform sampling baseline, the stratified sampling baseline, ABae, and \algname{} on the evaluation queries with and without a predicate. As shown in \autoref{tab:datasets}, our evaluation queries in the predicate setting cover a wide range of predicate positivity rates, with 37\% to 89\% of records matching the predicate. We swept the oracle budget from 500 to 5000 in increments of 500 and ran 1000 trials for each oracle budget. We executed \algname{} with its default hyperparameters ($K = 3$, $\alpha = 0.8$) on all datasets.



\begin{figure}
  \centering
  \includegraphics[width=\linewidth]{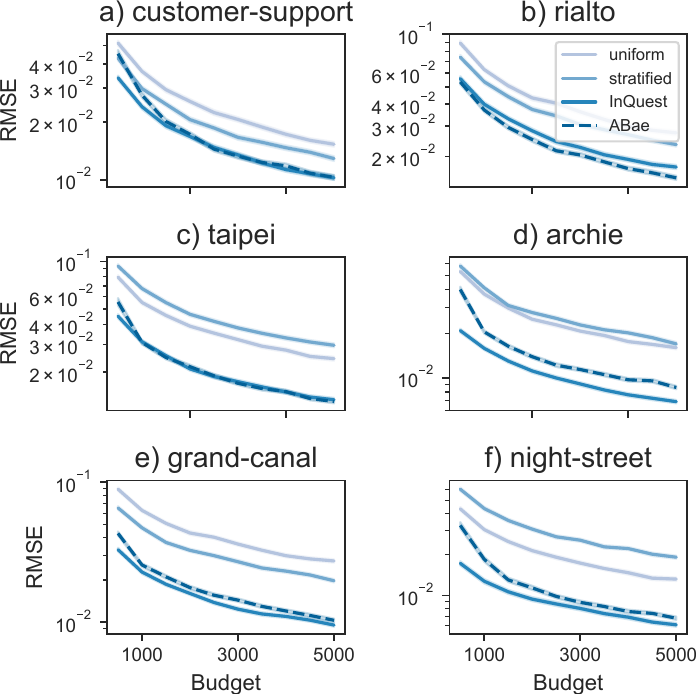}
  \caption{Sampling budget vs. median segment RMSE for baselines and \algname{} on the evaluation queries with no predicate (log scale). \algname{} outperforms the streaming baselines across all sampling budgets and datasets. \algname{} outperforms ABae on 70.0\% of oracle budgets across datasets.}
  \label{fig:no-pred-error}
\end{figure}

\begin{figure}
  \centering
  \includegraphics[width=\linewidth]{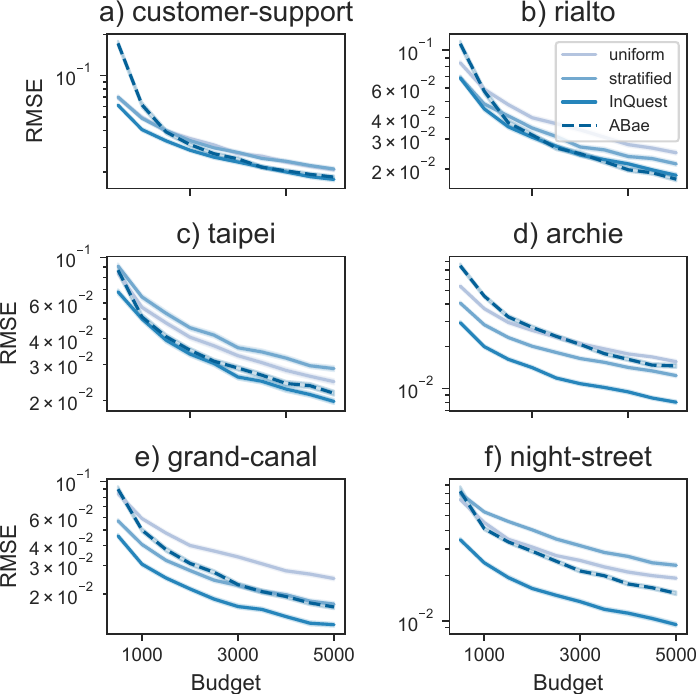}
  \caption{Sampling budget vs. median segment RMSE for baselines and \algname{} on the evaluation queries with a predicate (log scale). \algname{} outperforms the streaming baselines across all sampling budgets and datasets. \algname{} outperforms ABae on 90.0\% of oracle budgets across datasets.}
  \label{fig:pred-error}
\end{figure}

\autoref{fig:no-pred-error} shows \algname{} and the baselines' performance on the RMSE metric for the evaluation queries without a predicate. \algname{} outperforms streaming baselines on all sampling budgets across all datasets. \algname{} achieves as much as a 3.5x improvement on RMSE over streaming baselines at a fixed oracle budget, and can achieve the same RMSE with up to 5.0x fewer samples. \autoref{fig:pred-error} shows \algname{}'s results on queries with a predicate. Once again, \algname{} outperforms streaming baselines on all sampling budgets across all datasets. We demonstrate an improvement of up to 2.5x in RMSE at a fixed oracle budget over streaming baselines and are able to achieve the same error with up to 4.5x fewer samples.

We also compare \algname{} to ABae on the median segment RMSE metric. By default, ABae only returns an estimate for the entire query. We computed per-segment estimates by selecting the subset of ABae's oracle samples within each segment. We show that \algname{} outperforms ABae on 90.0\% and 70.0\% of oracle budgets across all datasets for queries with and without a predicate, respectively. We also compare \algname{} to ABae on the RMSE metric for the full query at the end of this subsection.

Finally, we quantified \algname{}'s performance relative to our baselines using a single error metric aggregated across all datasets. For each dataset and algorithm, we computed the mean of the median segment RMSEs over all 1000 trials at the given oracle budget. We then computed the geometric mean of these per-dataset average RMSEs to obtain a single aggregated error metric. We present these metrics in \autoref{tab:no-pred-all-results} and \autoref{tab:pred-all-results}. We can see that \algname{} achieves the lowest error metrics, outperforming streaming baselines by a factor of 1.32x-1.58x and 1.98x-2.05x across the entire range of oracle budgets for queries with and without a predicate, respectively. Furthermore, we demonstrate that \algname{} outperforms ABae by a factor of 1.04x-1.40x for queries without a predicate and by a factor of 1.26x-1.97x for queries with a predicate.




\begin{table}[t!]
  \caption{Summary of algorithm performance relative to streaming baselines and ABae in the no predicate case. RMSE errors are computed by taking the geometric mean of the average RMSE across all datasets at the specified budget.}
  \label{tab:no-pred-all-results}
  \begin{tabular}{lcccc}
    \toprule
    Algorithm & $NT = 500$ & $NT = 2500$ & $NT = 5000$ & All  \\
    \midrule
    $RMSE_{\textrm{uniform}^{\dag}}$ & .065 & .029 & .020 & .030 \\
    $RMSE_{\textrm{stratified}^{\ast}}$ & .064 & .029 & .020 & .030 \\
    $RMSE_{\textrm{ABae}^{\ddag}}$ & .044 & .015 & .010 & .016 \\
    $RMSE_{\algname{}}$ & \textbf{.032} & \textbf{.014} & \textbf{.0099} & \textbf{.015} \\
    \midrule
    $\textrm{Improvement}^{\dag}$ & 2.05x & 2.03x & 1.99x & 2.00x \\
    $\textrm{Improvement}^{\ast}$ & 2.01x & 2.02x & 1.98x & 2.00x \\
    $\textrm{Improvement}^{\ddag}$ & 1.40x & 1.05x & 1.04x & 1.10x \\
    \bottomrule
  \end{tabular}
\end{table}
\begin{table}[t!]
  \caption{Summary of algorithm performance relative to streaming baselines and ABae in predicate case. RMSE errors are computed by taking the geometric mean of the average RMSE across all datasets at the specified budget.}
  \label{tab:pred-all-results}
  \begin{tabular}{lcccc}
    \toprule
    Algorithm & $NT = 500$ & $NT = 2500$ & $NT = 5000$ & All  \\
    \midrule
    $RMSE_{\textrm{uniform}^{\dag}}$ & .072 & .032 & .021 & .033 \\
    $RMSE_{\textrm{stratified}^{\ast}}$ & .065 & .029 & .020 & .030 \\
    $RMSE_{\textrm{ABae}^{\ddag}}$ & .096 & .027 & .017 & .029 \\
    $RMSE_{\algname{}}$ & \textbf{.049} & \textbf{.020} & \textbf{.014} & \textbf{.021} \\
    \midrule
    $\textrm{Improvement}^{\dag}$ & 1.48x & 1.56x & 1.58x & 1.54x \\
    $\textrm{Improvement}^{\ast}$ & 1.32x & 1.43x & 1.48x & 1.42x \\
    $\textrm{Improvement}^{\ddag}$ & 1.97x & 1.32x & 1.26x & 1.37x \\
    \bottomrule
  \end{tabular}
\end{table}

\begin{figure}
  \centering
  \includegraphics[width=\linewidth]{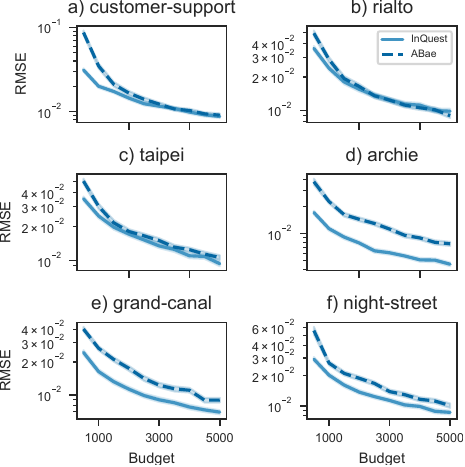}
  \caption{Sample budget vs. full query RMSE for \algname{} and ABae on the evaluation queries with a predicate (log scale).}
  \label{fig:pred-error-full-query}
\end{figure}

\minihead{Evaluating \algname{} and ABae on Full Query} We now compare \algname{} and ABae using the RMSE metric over the full evaluation query. This evaluation is useful for contextualizing \algname{}'s results relative to a state-of-the-art algorithm in the batch setting. We show the results for the queries with a predicate in \autoref{fig:pred-error-full-query}, and defer the figure for the queries without a predicate to an appendix \cite{techreport}.

When aggregating error results across all datasets we find that \algname{} outperforms ABae by a factor of 1.05x-1.41x on queries without a predicate, and by a factor of 1.18-1.83x for queries with a predicate. While one might expect ABae to provide an upper bound on \algname{}'s performance, \algname{} can benefit from its segmentation of certain streams. Specifically, if a stream is segmented over time such that $\sigma_{tk} < \sigma_{k}$ for enough segments $t \in [1, T]$, \algname{} can achieve more accurate estimates than a batch algorithm. To summarize, \algname{} can exploit the tendency in many real-world streams for proxy scores that are nearby in time to have similar values, which results in smaller $\sigma_{tk}$ and ultimately smaller errors when estimating $\mu$.




\subsection{Lesion and Sensitivity Analysis} \label{section:lesion-sensitivity}

\begin{figure}
  \centering
  \includegraphics[height=3.1in,width=\linewidth]{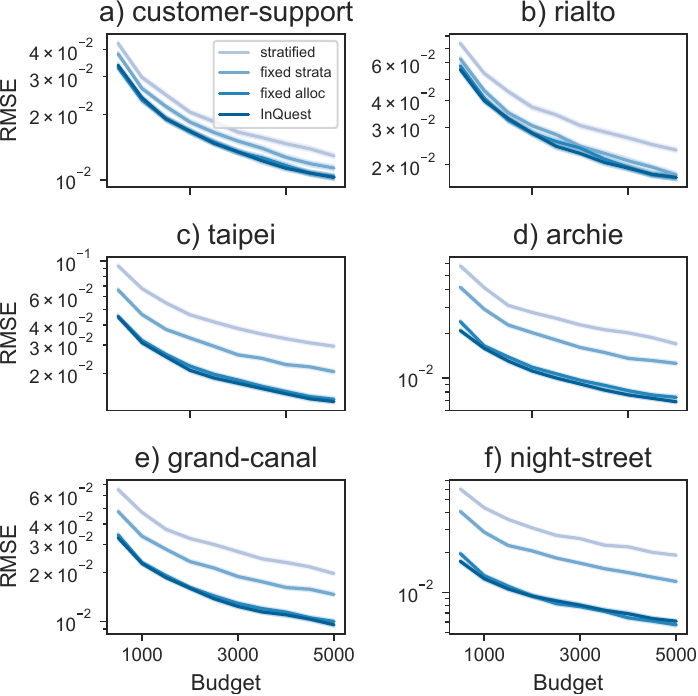}
  \caption{Lesion study in which we remove dynamic strata inference and dynamic sample allocation. As shown, both pieces of the algorithm are critical for achieving better performance across all datasets. All experiments were run on queries with no predicate.}
  \label{fig:lesion-study}
\end{figure}

\minihead{Lesion study}
We investigated whether all of \algname{}'s components were necessary for high performance. We performed a lesion study by executing (1) \algname{}, (2) \algname{} with dynamic strata inference but fixed sample allocations, (3) \algname{} with fixed strata but dynamic sample allocations, and (4) stratified sampling with a pilot segment. All experiments were run with 1000 trials on the evaluation queries with no predicate.

As shown in \autoref{fig:lesion-study}, both dynamic strata inference and dynamic sample allocations are important for achieving high performance. In particular, removing dynamic strata inference can severely limit \algname{}'s ability to avoid wasting samples on strata with few predicate matching records. While removing dynamic sample allocation does not significantly affect performance on some datasets (e.g., \texttt{grand-canal}), it is necessary for achieving high performance on others (e.g., \texttt{archie}, \texttt{rialto}, and \texttt{taipei}).

\minihead{Sensitivity analysis}
We further investigated whether \algname{}'s performance was sensitive to the setting of its key parameters. Specifically, we analyzed the sensitivity of \algname{} to the smoothing parameter $\alpha$ and to the length of its tumbling window. We ran \algname{} with a budget of 5000 samples for 1000 trials while varying $\alpha$ and the window length. All experiments were run on the evaluation queries with no predicate.

\begin{figure}[t!]
  \centering
  \includegraphics[height=1.1in,width=\linewidth]{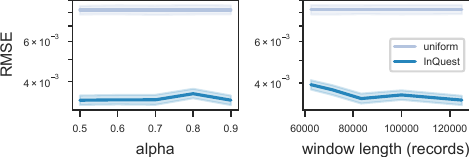}
  \caption{Sensitivity analysis of \algname{} to the smoothing parameter $\alpha$ and our tumbling window length on the \texttt{archie} dataset. As shown above, we can see that \algname{}'s RMSE is fairly stable w.r.t. changes in $\alpha$ and the window length.}
  \label{fig:sensitivity}
\end{figure}

\autoref{fig:sensitivity} shows \algname{}'s performance as a function of $\alpha$ and the window length on the \texttt{archie} dataset. \algname{}'s performance is relatively stable with respect to changes in $\alpha$ and the window length. We varied $\alpha \in [0.5, 0.9]$ in increments of 0.1 and we varied the window length such that the query contained $T \in [4,8]$ segments. We compared \algname{} to uniform sampling, which is invariant to these parameters. \algname{} outperforms uniform sampling on the RMSE metric on all datasets and settings of the $\alpha$ parameter and the window length. We defer the plots for the other datasets to our appendix for the sake of brevity.




\subsection{Cost Savings and Accuracy Improvements} We now examine how much \algname{} saves on cost---both in terms of time and dollars---relative to baselines. For each algorithm we measure its accuracy using the median segment RMSE. We compute the relative cost of running each algorithm as a function of its execution of its oracle and proxy models.

\begin{figure}
  \centering
  \includegraphics[height=1.1in,width=\linewidth]{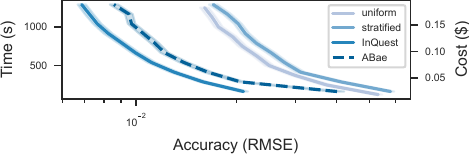}
  \caption{Time and cost in dollars as a function of accuracy for the \texttt{archie} dataset. At fixed accuracy, \algname{} achieves a speed up (and cost savings) of up to 5.8x over streaming baselines and up to 1.6x over ABae.}
  \label{fig:cost-savings-no-pred}
\end{figure}

We consider the case of our video datasets, where our oracle is a Mask R-CNN model that can process ~4 frames per second on an NVIDIA T4 GPU, and our proxy is a ResNet-18 model which can process ~12.6k frames per second \cite{kang2020jointly}. Using on-demand pricing from Amazon Web Services \cite{g42022pricing}, we assume the cost of running a single NVIDIA T4 GPU on a \texttt{g4dn.xlarge} is \$0.526 per hour. We compute the time (and associated cost) of running inference using our oracle and proxy models to achieve the stated accuracy. The results for the evaluation query on \texttt{archie} dataset without a predicate are presented in \autoref{fig:cost-savings-no-pred}. We can see that \algname{} outperforms all other algorithms on the \texttt{archie} dataset in terms of accuracy at fixed cost and cost at fixed accuracy. \algname{} achieves a speed up (and cost savings) of up to 5.8x over streaming baselines and up to 1.6x over ABae. \algname{} outperforms the baselines on the other datasets as well, achieving worst and best-case speedups of 1.5x-8.3x over streaming baselines and 0.8x-2.0x over ABae in the no predicate case, and of 1.0x-4.1x over streaming baselines and 0.9x-2.6x over ABae in the predicate case. We omit those plots for the sake of brevity and defer them to the appendix.

\subsection{Effect of Proxy Quality on Performance} 
\begin{figure}[t!]
  \centering
  \includegraphics[height=1.5in,width=\linewidth]{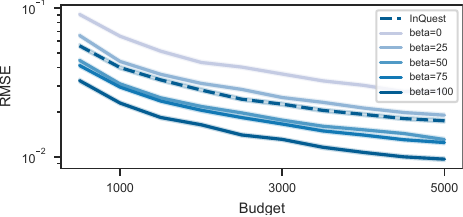}
  \caption{Proxy quality's effect on \algname{}'s performance on the \texttt{rialto} dataset. We plot \algname{}'s performance on the median segment RMSE metric as a function of $\beta$.}
  \label{fig:proxy-quality}
\end{figure}

In this subsection we examine how proxy score quality affects \algname{}'s performance on our evaluation datasets. We modified the proxy scores for our evaluation datasets by interpolating between the groundtruth statistic (i.e., perfect proxy information) and random noise. Specifically, for $\beta \in [0, 1]$ we computed:
\begin{equation}
    \textrm{proxy}_i = \beta \cdot \textrm{g}_i + (1 - \beta) \cdot \mathbb{U}(0,1)
\end{equation}
Where $\textrm{g}_i$ is the groundtruth statistic for the $i^{th}$ dataset record and $\mathbb{U}(0,1)$ is a sample drawn uniformly at random from the range [0, 1]. After computing the proxy values for each stream we normalized them to be in the range [0, 1]. We constructed datasets in this fashion for $\beta \in [0.0, 0.25, 0.50, 0.75, 1.0]$. In figure \autoref{fig:proxy-quality} we plot \algname{}'s performance on the \texttt{rialto} dataset as a function of $\beta$. We chose this dataset because its proxy's Pearson correlation coefficient is near the median for our datasets (see \autoref{tab:datasets}).

As shown in \autoref{fig:proxy-quality}, proxy quality can result in an orders-of-magnitude improvement of \algname{}'s performance. However, our current proxy scores are far from optimal, resulting in performances comparable to those with $\beta \in [0.25, 0.75]$. Thus, while we would expect ABae's performance to also improve with better proxies, we can confidently state that our performance relative to our uniform sampling baseline would greatly improve with better proxies.

\subsection{Adversarial Shifts in Stream Parameters}
We now investigate how one or more sudden shifts in the stream parameters $p_{tk}$, $\sigma_{tk}$, and $\mu_{tk}$ affects \algname{}'s performance.

\minihead{Dataset construction} We constructed streams by randomly inserting $n = [1, 2, \dots, 5]$ sudden shifts in the stream parameters. For each value of $n$ we generated 20 streams, thus creating a total of 100 synthetic datasets. To generate a stream, we began by sampling $n$ indices uniformly at random where we would suddenly shift the stream parameters. We then sampled our initial stream parameters $p_{1k}$, $\sigma_{1k}$, and $\mu_{1k}$ where: $p_{tk} \in [0, 1]$, $\sigma_{tk} \in [0, 3]$, and $(\mu_{t1}, \mu_{t2}, \mu_{t3}) \in ([0, 3], [3, 6], [6, 9])$.

For each value of $k \in [1, K]$ we generated a substream of samples using parameters ($p_{1k}$, $\sigma_{1k}$, $\mu_{1k}$). We then interleaved the samples from our $K$ substreams into our final synthetic streaming dataset until we reached the sample index for a sudden shift in parameters. At every such index, we resampled $p_{tk}$, $\sigma_{tk}$, and $\mu_{tk}$ for all $k \in [1, K]$ and continued constructing the streaming dataset with the new stream parameters in the same fashion. Finally, we computed synthetic proxy values by interpolating the groundtruth statistic in an identical fashion to our experiments in Section 5.5. For our synthetic datasets, we used $\beta=0.75$ to construct the proxies.



\begin{figure}[t!]
  \centering
  \includegraphics[height=1.1in,width=\linewidth]{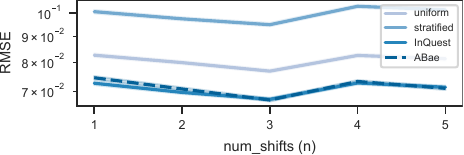}
  \caption{Analyzing the effect of shifts in the stream parameters on \algname{}'s performance. \algname{} outperforms streaming baselines by an average of 1.13x-1.42x on the median RMSE metric when evaluated on 100 synthetic datasets.}
  \label{fig:distribution-shift}
\end{figure}

While our theoretical analysis focused on \algname{}'s performance on stationary streams, by construction these synthetic datasets stress test \algname{}'s ability to adjust to sudden changes in dynamic streams. We evaluated our streaming baselines, ABae, and \algname{} for 1000 trials on all 100 synthetic datasets. We computed each algorithm's average median RMSE across all datasets for fixed values of $n$ and present the results in \autoref{fig:distribution-shift}.

\minihead{Results} \algname{} consistently outperforms our streaming baselines on these synthetic datasets, and performs comparably to ABae. \algname{} achieves 1.13x-1.42x improvement over our streaming baselines across the full range of $n \in [1, 5]$ distribution shifts. \algname{} also performs within 0.99x-1.03x of ABae. These results provide empirical evidence that \algname{} performs well even on non-stationary streams with multiple sudden shifts in the stream parameters.

%% file: tex/rel_work.tex
\section{Related Work}
We examine prior work in query processing as it relates to \algname{}. We first discuss prior methods for processing stream queries, particularly on video datasets. We then examine other AQP systems with a focus on DNN-based queries. Finally, we discuss literature related to proxy models and their use in query processing.

\minihead{Stream queries}
Prior work has focused on building systems to answer queries over (semi-)structured streams of data. Similar to \algname{}, these systems support continuous queries for real-time analytics, in which a user can submit a query for an indefinite period of time to compute a statistic of interest (e.g., the number of cars that pass over a sensor on the highway) \cite{katsifodimos2016flink, zaharia2013spark, arasu2003stream, madden2002fjording, chandrasekaran2002streaming}. However, because these systems are designed to work with structured data they do not address the use case where a DNN is needed to extract statistics of interest from the raw data stream. 


Recent work has focused on answering spatiotemporal queries over streaming video, including aggregation queries \cite{koudas2020video}, queries related to object co-occurrences \cite{chen2020evaluating}, and queries related to object interactions \cite{xarchakos2021querying}. Similar to \algname{}, these systems use cascades of filters to limit the execution of an oracle to a subset of the stream. However, these systems bake their filters into various layers of the oracle. This creates a tight-coupling between the design of the filters and the oracle and limits these systems to processing queries over video. In contrast, \algname{} decouples the proxy(s) from the oracle, thus enabling users to provide custom models which makes it easy for \algname{} to work across different modalities of data.


\minihead{AQP with DNN-based queries}
Recent work has focused on accelerating DNN-based queries over large unstructured datasets in the batch setting \cite{anderson2019tahoma, hsieh2018focus, kang2017noscope, kang2018blazeit, kang2020approximate, kang2021abae, moll2022exsample, miris2020bastani}. While these systems are similar to \algname{} in their use of DNNs to answer queries over large unstructured datasets, certain features of their designs make it difficult to adapt them to the streaming setting. For example, NoScope \cite{kang2017noscope} and Tahoma \cite{anderson2019tahoma} rely on drawing a representative sample from the full dataset before query submission in order to train and validate specialized DNNs and model cascades. ABae \cite{kang2021abae} and SUPG \cite{kang2020approximate} use sampling techniques over the entire dataset to optimize their oracle sampling strategies. ExSample \cite{moll2022exsample} takes the full dataset and splits it into chunks before query submission in order to perform Thompson sampling \cite{russo2017thompson} across these chunks. These systems would need to be modified substantially to work in the streaming setting, where the dataset is presented to the system in an online fashion.

\minihead{Proxies in query processing}
The use of proxy models for computing cheap approximations spans a variety of use cases in query processing. In certain video analytics systems \cite{canel2019scaling, koudas2020video}, proxies only compute binary predicates and they are all implemented in a single DNN (potentially at different layers). This is in contrast to our work, in which users provide proxies that can compute arbitrary statistics independent from the oracle. Systems such as NoScope, ABae, and SUPG \cite{kang2017noscope, kang2021abae, kang2020approximate} use proxies to estimate query predicates and thereby limit the execution of an expensive oracle to a subset of some large dataset. \algname{} uses proxies in a similar manner for processing queries with a predicate. For queries without a predicate \algname{} can use a proxy that computes any bounded real-valued estimate, but it will produce better results if the proxy estimate is correlated with the query's statistic of interest.

%% file: tex/conclusion.tex
\section{Conclusion}
In this work we proposed and analyzed \algname{}, a system for accelerating aggregation queries over unstructured streams of data with statistical guarantees on query accuracy. We demonstrated significant improvements over streaming and batch setting baselines on a set of real-world video and text datasets. We further showed that \algname{} is not sensitive to its parameter settings, that its major components are all crucial for its performance improvements, and that it is resilient to adversarial shifts in the ground-truth stream parameters. We performed a theoretical analysis and showed that \algname{}'s sample allocation converged to the optimal sample allocation and that its expected error converged to zero at quantitative rates. To the best of our knowledge, this is the first system designed for processing aggregation queries over streams of multiple modalities. Thus, \algname{} has the potential to be applied to a wide range of real-world problems, from processing queries over large networks of streaming video cameras to streams of social media posts.

%% file: main.bbl

\begin{thebibliography}{48}


\ifx \showCODEN    \undefined \def \showCODEN     #1{\unskip}     \fi
\ifx \showDOI      \undefined \def \showDOI       #1{#1}\fi
\ifx \showISBNx    \undefined \def \showISBNx     #1{\unskip}     \fi
\ifx \showISBNxiii \undefined \def \showISBNxiii  #1{\unskip}     \fi
\ifx \showISSN     \undefined \def \showISSN      #1{\unskip}     \fi
\ifx \showLCCN     \undefined \def \showLCCN      #1{\unskip}     \fi
\ifx \shownote     \undefined \def \shownote      #1{#1}          \fi
\ifx \showarticletitle \undefined \def \showarticletitle #1{#1}   \fi
\ifx \showURL      \undefined \def \showURL       {\relax}        \fi
\providecommand\bibfield[2]{#2}
\providecommand\bibinfo[2]{#2}
\providecommand\natexlab[1]{#1}
\providecommand\showeprint[2][]{arXiv:#2}

\bibitem[\protect\citeauthoryear{Aggarwal}{Aggarwal}{2006}]%
        {aggarwal2006biased}
\bibfield{author}{\bibinfo{person}{Charu~C Aggarwal}.}
  \bibinfo{year}{2006}\natexlab{}.
\newblock \showarticletitle{On biased reservoir sampling in the presence of
  stream evolution}. In \bibinfo{booktitle}{\emph{Proceedings of the 32nd
  international conference on Very large data bases}}.
  \bibinfo{pages}{607--618}.
\newblock


\bibitem[\protect\citeauthoryear{Al-Kateb, Lee, and Wang}{Al-Kateb
  et~al\mbox{.}}{2007}]%
        {al2007adaptive}
\bibfield{author}{\bibinfo{person}{Mohammed Al-Kateb},
  \bibinfo{person}{Byung~Suk Lee}, {and} \bibinfo{person}{X~Sean Wang}.}
  \bibinfo{year}{2007}\natexlab{}.
\newblock \showarticletitle{Adaptive-size reservoir sampling over data
  streams}. In \bibinfo{booktitle}{\emph{19th International Conference on
  Scientific and Statistical Database Management (SSDBM 2007)}}. IEEE,
  \bibinfo{pages}{22--22}.
\newblock


\bibitem[\protect\citeauthoryear{ALERTWildfire}{ALERTWildfire}{2022}]%
        {alertwildfire2022cameras}
\bibfield{author}{\bibinfo{person}{ALERTWildfire}.}
  \bibinfo{year}{2022}\natexlab{}.
\newblock \bibinfo{title}{AlertWildfire}.
\newblock
\newblock
\urldef\tempurl%
\url{https://www.alertwildfire.org/}
\showURL{%
Retrieved Dec. 28, 2022 from \tempurl}


\bibitem[\protect\citeauthoryear{Anderson, Cafarella, Ros, and
  Wenisch}{Anderson et~al\mbox{.}}{2019}]%
        {anderson2019tahoma}
\bibfield{author}{\bibinfo{person}{Michael~R. Anderson},
  \bibinfo{person}{Michael Cafarella}, \bibinfo{person}{German Ros}, {and}
  \bibinfo{person}{Thomas~F. Wenisch}.} \bibinfo{year}{2019}\natexlab{}.
\newblock \showarticletitle{Physical Representation-Based Predicate
  Optimization for a Visual Analytics Database}. In
  \bibinfo{booktitle}{\emph{2019 {IEEE} 35th International Conference on Data
  Engineering ({ICDE})}}. \bibinfo{publisher}{{IEEE}},
  \bibinfo{pages}{1466--1477}.
\newblock
\urldef\tempurl%
\url{https://doi.org/10.1109/icde.2019.00132}
\showDOI{\tempurl}


\bibitem[\protect\citeauthoryear{Arasu, Babcock, Babu, Datar, Ito, Nishizawa,
  Rosenstein, and Widom}{Arasu et~al\mbox{.}}{2003}]%
        {arasu2003stream}
\bibfield{author}{\bibinfo{person}{Arvind Arasu}, \bibinfo{person}{Brian
  Babcock}, \bibinfo{person}{Shivnath Babu}, \bibinfo{person}{Mayur Datar},
  \bibinfo{person}{Keith Ito}, \bibinfo{person}{Itaru Nishizawa},
  \bibinfo{person}{Justin Rosenstein}, {and} \bibinfo{person}{Jennifer Widom}.}
  \bibinfo{year}{2003}\natexlab{}.
\newblock \showarticletitle{STREAM: the stanford stream data manager
  (demonstration description)}. In \bibinfo{booktitle}{\emph{Proceedings of the
  2003 ACM SIGMOD international conference on Management of data}}.
  \bibinfo{pages}{665--665}.
\newblock


\bibitem[\protect\citeauthoryear{Babu and Widom}{Babu and Widom}{2001}]%
        {babu2001continuous}
\bibfield{author}{\bibinfo{person}{Shivnath Babu} {and}
  \bibinfo{person}{Jennifer Widom}.} \bibinfo{year}{2001}\natexlab{}.
\newblock \showarticletitle{Continuous queries over data streams}.
\newblock \bibinfo{journal}{\emph{ACM Sigmod Record}} \bibinfo{volume}{30},
  \bibinfo{number}{3} (\bibinfo{year}{2001}), \bibinfo{pages}{109--120}.
\newblock


\bibitem[\protect\citeauthoryear{Bastani, He, Balasingam, Gopalakrishnan,
  Alizadeh, Balakrishnan, Cafarella, Kraska, and Madden}{Bastani
  et~al\mbox{.}}{2020}]%
        {miris2020bastani}
\bibfield{author}{\bibinfo{person}{Favyen Bastani}, \bibinfo{person}{Songtao
  He}, \bibinfo{person}{Arjun Balasingam}, \bibinfo{person}{Karthik
  Gopalakrishnan}, \bibinfo{person}{Mohammad Alizadeh}, \bibinfo{person}{Hari
  Balakrishnan}, \bibinfo{person}{Michael Cafarella}, \bibinfo{person}{Tim
  Kraska}, {and} \bibinfo{person}{Sam Madden}.}
  \bibinfo{year}{2020}\natexlab{}.
\newblock \showarticletitle{MIRIS: Fast Object Track Queries in Video}. In
  \bibinfo{booktitle}{\emph{Proceedings of the 2020 ACM SIGMOD International
  Conference on Management of Data}} (Portland, OR, USA)
  \emph{(\bibinfo{series}{SIGMOD '20})}. \bibinfo{publisher}{Association for
  Computing Machinery}, \bibinfo{address}{New York, NY, USA},
  \bibinfo{pages}{1907–1921}.
\newblock
\showISBNx{9781450367356}
\urldef\tempurl%
\url{https://doi.org/10.1145/3318464.3389692}
\showDOI{\tempurl}


\bibitem[\protect\citeauthoryear{Bojanowski, Grave, Joulin, and
  Mikolov}{Bojanowski et~al\mbox{.}}{2017}]%
        {bojanowski2016enriching}
\bibfield{author}{\bibinfo{person}{Piotr Bojanowski}, \bibinfo{person}{Edouard
  Grave}, \bibinfo{person}{Armand Joulin}, {and} \bibinfo{person}{Tom{\'{a}}s
  Mikolov}.} \bibinfo{year}{2017}\natexlab{}.
\newblock \showarticletitle{Enriching Word Vectors with Subword Information}.
  In \bibinfo{booktitle}{\emph{Transactions of the Association for
  Computational Linguistics}}, Vol.~\bibinfo{volume}{5}.
  \bibinfo{pages}{135--146}.
\newblock
\showeprint[arXiv]{1607.04606}
\urldef\tempurl%
\url{http://arxiv.org/abs/1607.04606}
\showURL{%
\tempurl}


\bibitem[\protect\citeauthoryear{Braverman and Ostrovsky}{Braverman and
  Ostrovsky}{2013}]%
        {braverman2014generalizing}
\bibfield{author}{\bibinfo{person}{Vladimir Braverman} {and}
  \bibinfo{person}{Rafail Ostrovsky}.} \bibinfo{year}{2013}\natexlab{}.
\newblock \showarticletitle{Generalizing the layering method of indyk and
  woodruff: Recursive sketches for frequency-based vectors on streams}. In
  \bibinfo{booktitle}{\emph{Approximation, Randomization, and Combinatorial
  Optimization. Algorithms and Techniques.}} \bibinfo{publisher}{Springer},
  \bibinfo{pages}{58--70}.
\newblock


\bibitem[\protect\citeauthoryear{Browne}{Browne}{1995}]%
        {browne1995use}
\bibfield{author}{\bibinfo{person}{Richard~H Browne}.}
  \bibinfo{year}{1995}\natexlab{}.
\newblock \showarticletitle{On the use of a pilot sample for sample size
  determination}.
\newblock \bibinfo{journal}{\emph{Statistics in medicine}}
  \bibinfo{volume}{14}, \bibinfo{number}{17} (\bibinfo{year}{1995}),
  \bibinfo{pages}{1933--1940}.
\newblock


\bibitem[\protect\citeauthoryear{Canel, Kim, Zhou, Li, Lim, Andersen, Kaminsky,
  and Dulloor}{Canel et~al\mbox{.}}{2019}]%
        {canel2019scaling}
\bibfield{author}{\bibinfo{person}{Christopher Canel}, \bibinfo{person}{Thomas
  Kim}, \bibinfo{person}{Giulio Zhou}, \bibinfo{person}{Conglong Li},
  \bibinfo{person}{Hyeontaek Lim}, \bibinfo{person}{David~G. Andersen},
  \bibinfo{person}{Michael Kaminsky}, {and} \bibinfo{person}{Subramanya~R.
  Dulloor}.} \bibinfo{year}{2019}\natexlab{}.
\newblock \showarticletitle{Scaling Video Analytics on Constrained Edge Nodes}.
  In \bibinfo{booktitle}{\emph{Proceedings of the 2nd SysML Conference}}.
  \bibinfo{address}{Palo Alto, CA, USA}, 12 pages.
\newblock
\showeprint[arXiv]{1905.13536}
\urldef\tempurl%
\url{http://arxiv.org/abs/1905.13536}
\showURL{%
\tempurl}


\bibitem[\protect\citeauthoryear{Chandrasekaran and Franklin}{Chandrasekaran
  and Franklin}{2002}]%
        {chandrasekaran2002streaming}
\bibfield{author}{\bibinfo{person}{Sirish Chandrasekaran} {and}
  \bibinfo{person}{Michael~J Franklin}.} \bibinfo{year}{2002}\natexlab{}.
\newblock \showarticletitle{Streaming queries over streaming data}. In
  \bibinfo{booktitle}{\emph{VLDB'02: Proceedings of the 28th International
  Conference on Very Large Databases}}. Elsevier, \bibinfo{pages}{203--214}.
\newblock


\bibitem[\protect\citeauthoryear{Chen, Yu, Koudas, and Yu}{Chen
  et~al\mbox{.}}{2021}]%
        {chen2020evaluating}
\bibfield{author}{\bibinfo{person}{Yueting Chen}, \bibinfo{person}{Xiaohui Yu},
  \bibinfo{person}{Nick Koudas}, {and} \bibinfo{person}{Ziqiang Yu}.}
  \bibinfo{year}{2021}\natexlab{}.
\newblock \showarticletitle{Evaluating Temporal Queries Over Video Feeds}. In
  \bibinfo{booktitle}{\emph{Proceedings of the 2021 International Conference on
  Management of Data}} (Virtual Event, China) \emph{(\bibinfo{series}{SIGMOD
  '21})}. \bibinfo{publisher}{Association for Computing Machinery},
  \bibinfo{address}{New York, NY, USA}, \bibinfo{pages}{287–299}.
\newblock
\showISBNx{9781450383431}
\urldef\tempurl%
\url{https://doi.org/10.1145/3448016.3452803}
\showDOI{\tempurl}


\bibitem[\protect\citeauthoryear{Devlin, Chang, Lee, and Toutanova}{Devlin
  et~al\mbox{.}}{2019}]%
        {devlin2018bert}
\bibfield{author}{\bibinfo{person}{Jacob Devlin}, \bibinfo{person}{Ming{-}Wei
  Chang}, \bibinfo{person}{Kenton Lee}, {and} \bibinfo{person}{Kristina
  Toutanova}.} \bibinfo{year}{2019}\natexlab{}.
\newblock \showarticletitle{{BERT:} Pre-training of Deep Bidirectional
  Transformers for Language Understanding}. In
  \bibinfo{booktitle}{\emph{Proceedings of the 2019 Conference of the North
  American Chapter of the Association for Computational Linguistics: Human
  Language Technologies, {NAACL-HLT} 2019, Minneapolis, MN, USA, June 2-7,
  2019, Volume 1 (Long and Short Papers)}},
  \bibfield{editor}{\bibinfo{person}{Jill Burstein}, \bibinfo{person}{Christy
  Doran}, {and} \bibinfo{person}{Thamar Solorio}} (Eds.).
  \bibinfo{publisher}{Association for Computational Linguistics},
  \bibinfo{pages}{4171--4186}.
\newblock
\urldef\tempurl%
\url{https://doi.org/10.18653/v1/n19-1423}
\showDOI{\tempurl}


\bibitem[\protect\citeauthoryear{EC2}{EC2}{2023}]%
        {g42022pricing}
\bibfield{author}{\bibinfo{person}{AWS EC2}.} \bibinfo{year}{2023}\natexlab{}.
\newblock \bibinfo{title}{G4 On-Demand Pricing}.
\newblock
\newblock
\urldef\tempurl%
\url{https://aws.amazon.com/ec2/instance-types/g4/}
\showURL{%
Retrieved Apr. 11, 2023 from \tempurl}


\bibitem[\protect\citeauthoryear{Flink}{Flink}{2023}]%
        {flink2023syntax}
\bibfield{author}{\bibinfo{person}{Apache Flink}.}
  \bibinfo{year}{2023}\natexlab{}.
\newblock \bibinfo{title}{Tumbling Windows}.
\newblock
\newblock
\urldef\tempurl%
\url{https://nightlies.apache.org/flink/flink-docs-master/docs/dev/datastream/operators/windows/#tumbling-windows}
\showURL{%
Retrieved May 13, 2023 from \tempurl}


\bibitem[\protect\citeauthoryear{Fu, Crichton, Hong, Yao, Zhang, Truong,
  Narayan, Agrawala, R{\'{e}}, and Fatahalian}{Fu et~al\mbox{.}}{2019}]%
        {fu2019rekall}
\bibfield{author}{\bibinfo{person}{Daniel~Y. Fu}, \bibinfo{person}{Will
  Crichton}, \bibinfo{person}{James Hong}, \bibinfo{person}{Xinwei Yao},
  \bibinfo{person}{Haotian Zhang}, \bibinfo{person}{Anh Truong},
  \bibinfo{person}{Avanika Narayan}, \bibinfo{person}{Maneesh Agrawala},
  \bibinfo{person}{Christopher R{\'{e}}}, {and} \bibinfo{person}{Kayvon
  Fatahalian}.} \bibinfo{year}{2019}\natexlab{}.
\newblock \showarticletitle{Rekall: Specifying Video Events using Compositions
  of Spatiotemporal Labels}. In \bibinfo{booktitle}{\emph{SOSP 2019 Workshop on
  AI Systems}}. 16 pages.
\newblock
\showeprint[arXiv]{1910.02993}
\urldef\tempurl%
\url{http://arxiv.org/abs/1910.02993}
\showURL{%
\tempurl}


\bibitem[\protect\citeauthoryear{He, Gkioxari, Dollár, and Girshick}{He
  et~al\mbox{.}}{2017}]%
        {he2017maskrcnn}
\bibfield{author}{\bibinfo{person}{Kaiming He}, \bibinfo{person}{Georgia
  Gkioxari}, \bibinfo{person}{Piotr Dollár}, {and} \bibinfo{person}{Ross
  Girshick}.} \bibinfo{year}{2017}\natexlab{}.
\newblock \showarticletitle{Mask R-CNN}. In \bibinfo{booktitle}{\emph{2017 IEEE
  International Conference on Computer Vision (ICCV)}}.
  \bibinfo{pages}{2980--2988}.
\newblock
\urldef\tempurl%
\url{https://doi.org/10.1109/ICCV.2017.322}
\showDOI{\tempurl}


\bibitem[\protect\citeauthoryear{He, Zhang, Ren, and Sun}{He
  et~al\mbox{.}}{2016}]%
        {he2015resnet}
\bibfield{author}{\bibinfo{person}{Kaiming He}, \bibinfo{person}{Xiangyu
  Zhang}, \bibinfo{person}{Shaoqing Ren}, {and} \bibinfo{person}{Jian Sun}.}
  \bibinfo{year}{2016}\natexlab{}.
\newblock \showarticletitle{Deep Residual Learning for Image Recognition}. In
  \bibinfo{booktitle}{\emph{2016 IEEE Conference on Computer Vision and Pattern
  Recognition (CVPR)}}. \bibinfo{pages}{770--778}.
\newblock
\urldef\tempurl%
\url{https://doi.org/10.1109/CVPR.2016.90}
\showDOI{\tempurl}


\bibitem[\protect\citeauthoryear{Hong, Crichton, Zhang, Fu, Ritchie,
  Barenholtz, Hannel, Yao, Murray, Moriba, Agrawala, and Fatahalian}{Hong
  et~al\mbox{.}}{2021}]%
        {hong2020analyzing}
\bibfield{author}{\bibinfo{person}{James Hong}, \bibinfo{person}{Will
  Crichton}, \bibinfo{person}{Haotian Zhang}, \bibinfo{person}{Daniel~Y. Fu},
  \bibinfo{person}{Jacob Ritchie}, \bibinfo{person}{Jeremy Barenholtz},
  \bibinfo{person}{Ben Hannel}, \bibinfo{person}{Xinwei Yao},
  \bibinfo{person}{Michaela Murray}, \bibinfo{person}{Geraldine Moriba},
  \bibinfo{person}{Maneesh Agrawala}, {and} \bibinfo{person}{Kayvon
  Fatahalian}.} \bibinfo{year}{2021}\natexlab{}.
\newblock \showarticletitle{Analysis of Faces in a Decade of US Cable TV News}.
  In \bibinfo{booktitle}{\emph{Proceedings of the 27th ACM SIGKDD Conference on
  Knowledge Discovery \& Data Mining}} (Virtual Event, Singapore)
  \emph{(\bibinfo{series}{KDD '21})}. \bibinfo{publisher}{Association for
  Computing Machinery}, \bibinfo{address}{New York, NY, USA},
  \bibinfo{pages}{3011–3021}.
\newblock
\showISBNx{9781450383325}
\urldef\tempurl%
\url{https://doi.org/10.1145/3447548.3467134}
\showDOI{\tempurl}


\bibitem[\protect\citeauthoryear{Hsieh, Ananthanarayanan, Bodik, Venkataraman,
  Bahl, Philipose, Gibbons, and Mutlu}{Hsieh et~al\mbox{.}}{2018}]%
        {hsieh2018focus}
\bibfield{author}{\bibinfo{person}{Kevin Hsieh}, \bibinfo{person}{Ganesh
  Ananthanarayanan}, \bibinfo{person}{Peter Bodik}, \bibinfo{person}{Shivaram
  Venkataraman}, \bibinfo{person}{Paramvir Bahl}, \bibinfo{person}{Matthai
  Philipose}, \bibinfo{person}{Phillip~B. Gibbons}, {and} \bibinfo{person}{Onur
  Mutlu}.} \bibinfo{year}{2018}\natexlab{}.
\newblock \showarticletitle{Focus: Querying Large Video Datasets with Low
  Latency and Low Cost}. In \bibinfo{booktitle}{\emph{13th USENIX Symposium on
  Operating Systems Design and Implementation (OSDI 18)}}.
  \bibinfo{publisher}{USENIX Association}, \bibinfo{address}{Carlsbad, CA},
  \bibinfo{pages}{269--286}.
\newblock
\showISBNx{978-1-939133-08-3}
\urldef\tempurl%
\url{https://www.usenix.org/conference/osdi18/presentation/hsieh}
\showURL{%
\tempurl}


\bibitem[\protect\citeauthoryear{HuggingFace}{HuggingFace}{2022}]%
        {huggingface2022model}
\bibfield{author}{\bibinfo{person}{HuggingFace}.}
  \bibinfo{year}{2022}\natexlab{}.
\newblock \bibinfo{title}{Twitter-roBERTa-base for Sentiment Analysis}.
\newblock
\newblock
\urldef\tempurl%
\url{https://huggingface.co/cardiffnlp/twitter-roberta-base-sentiment-latest}
\showURL{%
Retrieved Dec. 29, 2022 from \tempurl}


\bibitem[\protect\citeauthoryear{Kaggle}{Kaggle}{2022}]%
        {kaggle2022dataset}
\bibfield{author}{\bibinfo{person}{Kaggle}.} \bibinfo{year}{2022}\natexlab{}.
\newblock \bibinfo{title}{Customer Support on Twitter}.
\newblock
\newblock
\urldef\tempurl%
\url{https://www.kaggle.com/datasets/thoughtvector/customer-support-on-twitter}
\showURL{%
Retrieved Dec. 29, 2022 from \tempurl}


\bibitem[\protect\citeauthoryear{Kang, Bailis, and Zaharia}{Kang
  et~al\mbox{.}}{2019}]%
        {kang2018blazeit}
\bibfield{author}{\bibinfo{person}{Daniel Kang}, \bibinfo{person}{Peter
  Bailis}, {and} \bibinfo{person}{Matei Zaharia}.}
  \bibinfo{year}{2019}\natexlab{}.
\newblock \showarticletitle{BlazeIt: Optimizing Declarative Aggregation and
  Limit Queries for Neural Network-Based Video Analytics}.
\newblock \bibinfo{journal}{\emph{Proc. VLDB Endow.}} \bibinfo{volume}{13},
  \bibinfo{number}{4} (\bibinfo{date}{dec} \bibinfo{year}{2019}),
  \bibinfo{pages}{533–546}.
\newblock
\showISSN{2150-8097}
\urldef\tempurl%
\url{https://doi.org/10.14778/3372716.3372725}
\showDOI{\tempurl}


\bibitem[\protect\citeauthoryear{Kang, Emmons, Abuzaid, Bailis, and
  Zaharia}{Kang et~al\mbox{.}}{2017}]%
        {kang2017noscope}
\bibfield{author}{\bibinfo{person}{Daniel Kang}, \bibinfo{person}{John Emmons},
  \bibinfo{person}{Firas Abuzaid}, \bibinfo{person}{Peter Bailis}, {and}
  \bibinfo{person}{Matei Zaharia}.} \bibinfo{year}{2017}\natexlab{}.
\newblock \showarticletitle{NoScope: Optimizing Neural Network Queries over
  Video at Scale}.
\newblock \bibinfo{journal}{\emph{Proc. VLDB Endow.}} \bibinfo{volume}{10},
  \bibinfo{number}{11} (\bibinfo{date}{aug} \bibinfo{year}{2017}),
  \bibinfo{pages}{1586–1597}.
\newblock
\showISSN{2150-8097}
\urldef\tempurl%
\url{https://doi.org/10.14778/3137628.3137664}
\showDOI{\tempurl}


\bibitem[\protect\citeauthoryear{Kang, Gan, Bailis, Hashimoto, and
  Zaharia}{Kang et~al\mbox{.}}{2020a}]%
        {kang2020approximate}
\bibfield{author}{\bibinfo{person}{Daniel Kang}, \bibinfo{person}{Edward Gan},
  \bibinfo{person}{Peter Bailis}, \bibinfo{person}{Tatsunori Hashimoto}, {and}
  \bibinfo{person}{Matei Zaharia}.} \bibinfo{year}{2020}\natexlab{a}.
\newblock \showarticletitle{Approximate Selection with Guarantees Using
  Proxies}.
\newblock \bibinfo{journal}{\emph{Proc. VLDB Endow.}} \bibinfo{volume}{13},
  \bibinfo{number}{12} (\bibinfo{date}{jul} \bibinfo{year}{2020}),
  \bibinfo{pages}{1990–2003}.
\newblock
\showISSN{2150-8097}
\urldef\tempurl%
\url{https://doi.org/10.14778/3407790.3407804}
\showDOI{\tempurl}


\bibitem[\protect\citeauthoryear{Kang, Guibas, Bailis, Hashimoto, Sun, and
  Zaharia}{Kang et~al\mbox{.}}{2021a}]%
        {kang2021abae}
\bibfield{author}{\bibinfo{person}{Daniel Kang}, \bibinfo{person}{John Guibas},
  \bibinfo{person}{Peter Bailis}, \bibinfo{person}{Tatsunori Hashimoto},
  \bibinfo{person}{Yi Sun}, {and} \bibinfo{person}{Matei Zaharia}.}
  \bibinfo{year}{2021}\natexlab{a}.
\newblock \showarticletitle{Accelerating Approximate Aggregation Queries with
  Expensive Predicates}.
\newblock \bibinfo{journal}{\emph{Proc. VLDB Endow.}} \bibinfo{volume}{14},
  \bibinfo{number}{11} (\bibinfo{date}{jul} \bibinfo{year}{2021}),
  \bibinfo{pages}{2341–2354}.
\newblock
\showISSN{2150-8097}
\urldef\tempurl%
\url{https://doi.org/10.14778/3476249.3476285}
\showDOI{\tempurl}


\bibitem[\protect\citeauthoryear{Kang, Guibas, Bailis, Hashimoto, Sun, and
  Zaharia}{Kang et~al\mbox{.}}{2021b}]%
        {abaetechreport}
\bibfield{author}{\bibinfo{person}{Daniel Kang}, \bibinfo{person}{John Guibas},
  \bibinfo{person}{Peter Bailis}, \bibinfo{person}{Tatsunori Hashimoto},
  \bibinfo{person}{Yi Sun}, {and} \bibinfo{person}{Matei Zaharia}.}
  \bibinfo{year}{2021}\natexlab{b}.
\newblock \bibinfo{title}{Proof: Accelerating Approximate Aggregation Queries
  with Expensive Predicates}.
\newblock
\newblock
\urldef\tempurl%
\url{https://ddkang.github.io/papers/2021/abae-tech-report.pdf}
\showURL{%
Retrieved Dec. 30, 2022 from \tempurl}


\bibitem[\protect\citeauthoryear{Kang, Guibas, Bailis, Hashimoto, and
  Zaharia}{Kang et~al\mbox{.}}{2022}]%
        {kang2020tasti}
\bibfield{author}{\bibinfo{person}{Daniel Kang}, \bibinfo{person}{John Guibas},
  \bibinfo{person}{Peter~D. Bailis}, \bibinfo{person}{Tatsunori Hashimoto},
  {and} \bibinfo{person}{Matei Zaharia}.} \bibinfo{year}{2022}\natexlab{}.
\newblock \showarticletitle{TASTI: Semantic Indexes for Machine Learning-Based
  Queries over Unstructured Data}. In \bibinfo{booktitle}{\emph{Proceedings of
  the 2022 International Conference on Management of Data}} (Philadelphia, PA,
  USA) \emph{(\bibinfo{series}{SIGMOD '22})}. \bibinfo{publisher}{Association
  for Computing Machinery}, \bibinfo{address}{New York, NY, USA},
  \bibinfo{pages}{1934–1947}.
\newblock
\showISBNx{9781450392495}
\urldef\tempurl%
\url{https://doi.org/10.1145/3514221.3517897}
\showDOI{\tempurl}


\bibitem[\protect\citeauthoryear{Kang, Mathur, Veeramacheneni, Bailis, and
  Zaharia}{Kang et~al\mbox{.}}{2020b}]%
        {kang2020jointly}
\bibfield{author}{\bibinfo{person}{Daniel Kang}, \bibinfo{person}{Ankit
  Mathur}, \bibinfo{person}{Teja Veeramacheneni}, \bibinfo{person}{Peter
  Bailis}, {and} \bibinfo{person}{Matei Zaharia}.}
  \bibinfo{year}{2020}\natexlab{b}.
\newblock \showarticletitle{Jointly Optimizing Preprocessing and Inference for
  DNN-Based Visual Analytics}.
\newblock \bibinfo{journal}{\emph{Proc. VLDB Endow.}} \bibinfo{volume}{14},
  \bibinfo{number}{2} (\bibinfo{date}{oct} \bibinfo{year}{2020}),
  \bibinfo{pages}{87–100}.
\newblock
\showISSN{2150-8097}
\urldef\tempurl%
\url{https://doi.org/10.14778/3425879.3425881}
\showDOI{\tempurl}


\bibitem[\protect\citeauthoryear{Katsifodimos and Schelter}{Katsifodimos and
  Schelter}{2016}]%
        {katsifodimos2016flink}
\bibfield{author}{\bibinfo{person}{Asterios Katsifodimos} {and}
  \bibinfo{person}{Sebastian Schelter}.} \bibinfo{year}{2016}\natexlab{}.
\newblock \showarticletitle{Apache Flink: Stream Analytics at Scale}. In
  \bibinfo{booktitle}{\emph{2016 IEEE International Conference on Cloud
  Engineering Workshop (IC2EW)}}. \bibinfo{pages}{193--193}.
\newblock
\urldef\tempurl%
\url{https://doi.org/10.1109/IC2EW.2016.56}
\showDOI{\tempurl}


\bibitem[\protect\citeauthoryear{Koudas, Li, and Xarchakos}{Koudas
  et~al\mbox{.}}{2022}]%
        {koudas2020video}
\bibfield{author}{\bibinfo{person}{Nick Koudas}, \bibinfo{person}{Raymond Li},
  {and} \bibinfo{person}{Ioannis Xarchakos}.} \bibinfo{year}{2022}\natexlab{}.
\newblock \showarticletitle{Video Monitoring Queries}.
\newblock \bibinfo{journal}{\emph{IEEE Trans. on Knowl. and Data Eng.}}
  \bibinfo{volume}{34}, \bibinfo{number}{10} (\bibinfo{date}{oct}
  \bibinfo{year}{2022}), \bibinfo{pages}{5023–5036}.
\newblock
\showISSN{1041-4347}
\urldef\tempurl%
\url{https://doi.org/10.1109/TKDE.2020.3048606}
\showDOI{\tempurl}


\bibitem[\protect\citeauthoryear{Loureiro, Barbieri, Neves, Espinosa~Anke, and
  Camacho-collados}{Loureiro et~al\mbox{.}}{2022}]%
        {loureiro2022timelms}
\bibfield{author}{\bibinfo{person}{Daniel Loureiro}, \bibinfo{person}{Francesco
  Barbieri}, \bibinfo{person}{Leonardo Neves}, \bibinfo{person}{Luis
  Espinosa~Anke}, {and} \bibinfo{person}{Jose Camacho-collados}.}
  \bibinfo{year}{2022}\natexlab{}.
\newblock \showarticletitle{{T}ime{LM}s: Diachronic Language Models from
  {T}witter}. In \bibinfo{booktitle}{\emph{Proceedings of the 60th Annual
  Meeting of the Association for Computational Linguistics: System
  Demonstrations}}. \bibinfo{publisher}{Association for Computational
  Linguistics}, \bibinfo{address}{Dublin, Ireland}, \bibinfo{pages}{251--260}.
\newblock
\urldef\tempurl%
\url{https://doi.org/10.18653/v1/2022.acl-demo.25}
\showDOI{\tempurl}


\bibitem[\protect\citeauthoryear{Madden and Franklin}{Madden and
  Franklin}{2002}]%
        {madden2002fjording}
\bibfield{author}{\bibinfo{person}{Samuel Madden} {and}
  \bibinfo{person}{Michael~J Franklin}.} \bibinfo{year}{2002}\natexlab{}.
\newblock \showarticletitle{Fjording the stream: An architecture for queries
  over streaming sensor data}. In \bibinfo{booktitle}{\emph{Proceedings 18th
  International Conference on Data Engineering}}. IEEE,
  \bibinfo{pages}{555--566}.
\newblock


\bibitem[\protect\citeauthoryear{Moll, Bastani, Madden, Stonebraker, Gadepally,
  and Kraska}{Moll et~al\mbox{.}}{2022}]%
        {moll2022exsample}
\bibfield{author}{\bibinfo{person}{Oscar Moll}, \bibinfo{person}{Favyen
  Bastani}, \bibinfo{person}{Sam Madden}, \bibinfo{person}{Mike Stonebraker},
  \bibinfo{person}{Vijay Gadepally}, {and} \bibinfo{person}{Tim Kraska}.}
  \bibinfo{year}{2022}\natexlab{}.
\newblock \showarticletitle{ExSample: Efficient Searches on Video Repositories
  through Adaptive Sampling}. In \bibinfo{booktitle}{\emph{2022 IEEE 38th
  International Conference on Data Engineering (ICDE)}}.
  \bibinfo{pages}{2956--2968}.
\newblock
\urldef\tempurl%
\url{https://doi.org/10.1109/ICDE53745.2022.00266}
\showDOI{\tempurl}


\bibitem[\protect\citeauthoryear{of~Science: Department~of
  Statistics}{of~Science: Department~of Statistics}{2023}]%
        {optimalallocation}
\bibfield{author}{\bibinfo{person}{Penn State Eberly~College of~Science:
  Department~of Statistics}.} \bibinfo{year}{2023}\natexlab{}.
\newblock \bibinfo{title}{Lesson 6: Stratified Sampling}.
\newblock
\newblock
\urldef\tempurl%
\url{https://online.stat.psu.edu/stat506/book/export/html/655}
\showURL{%
Retrieved Jul. 17, 2023 from \tempurl}


\bibitem[\protect\citeauthoryear{Owen and Zhou}{Owen and Zhou}{2000}]%
        {owen2000safe}
\bibfield{author}{\bibinfo{person}{Art Owen} {and} \bibinfo{person}{Yi Zhou}.}
  \bibinfo{year}{2000}\natexlab{}.
\newblock \showarticletitle{Safe and effective importance sampling}.
\newblock \bibinfo{journal}{\emph{J. Amer. Statist. Assoc.}}
  \bibinfo{volume}{95}, \bibinfo{number}{449} (\bibinfo{year}{2000}),
  \bibinfo{pages}{135--143}.
\newblock


\bibitem[\protect\citeauthoryear{Parsons}{Parsons}{2014}]%
        {parsons2014stratified}
\bibfield{author}{\bibinfo{person}{Van~L Parsons}.}
  \bibinfo{year}{2014}\natexlab{}.
\newblock \showarticletitle{Stratified sampling}.
\newblock \bibinfo{journal}{\emph{Wiley StatsRef: Statistics Reference Online}}
  (\bibinfo{year}{2014}), \bibinfo{pages}{1--11}.
\newblock


\bibitem[\protect\citeauthoryear{Piatetsky-Shapiro and
  Connell}{Piatetsky-Shapiro and Connell}{1984}]%
        {piatestsky-shapiro1984accurate}
\bibfield{author}{\bibinfo{person}{Gregory Piatetsky-Shapiro} {and}
  \bibinfo{person}{Charles Connell}.} \bibinfo{year}{1984}\natexlab{}.
\newblock \showarticletitle{Accurate Estimation of the Number of Tuples
  Satisfying a Condition}.
\newblock \bibinfo{journal}{\emph{SIGMOD Rec.}} \bibinfo{volume}{14},
  \bibinfo{number}{2} (\bibinfo{date}{jun} \bibinfo{year}{1984}),
  \bibinfo{pages}{256–276}.
\newblock
\showISSN{0163-5808}
\urldef\tempurl%
\url{https://doi.org/10.1145/971697.602294}
\showDOI{\tempurl}


\bibitem[\protect\citeauthoryear{Poms, Crichton, Hanrahan, and Fatahalian}{Poms
  et~al\mbox{.}}{2018}]%
        {poms2018scanner}
\bibfield{author}{\bibinfo{person}{Alex Poms}, \bibinfo{person}{Will Crichton},
  \bibinfo{person}{Pat Hanrahan}, {and} \bibinfo{person}{Kayvon Fatahalian}.}
  \bibinfo{year}{2018}\natexlab{}.
\newblock \showarticletitle{Scanner: Efficient Video Analysis at Scale}.
\newblock \bibinfo{journal}{\emph{ACM Trans. Graph.}} \bibinfo{volume}{37},
  \bibinfo{number}{4}, Article \bibinfo{articleno}{138} (\bibinfo{date}{jul}
  \bibinfo{year}{2018}), \bibinfo{numpages}{13}~pages.
\newblock
\showISSN{0730-0301}
\urldef\tempurl%
\url{https://doi.org/10.1145/3197517.3201394}
\showDOI{\tempurl}


\bibitem[\protect\citeauthoryear{Russo, Van~Roy, Kazerouni, Osband, and
  Wen}{Russo et~al\mbox{.}}{2018}]%
        {russo2017thompson}
\bibfield{author}{\bibinfo{person}{Daniel~J. Russo}, \bibinfo{person}{Benjamin
  Van~Roy}, \bibinfo{person}{Abbas Kazerouni}, \bibinfo{person}{Ian Osband},
  {and} \bibinfo{person}{Zheng Wen}.} \bibinfo{year}{2018}\natexlab{}.
\newblock \showarticletitle{A Tutorial on Thompson Sampling}.
\newblock \bibinfo{journal}{\emph{Found. Trends Mach. Learn.}}
  \bibinfo{volume}{11}, \bibinfo{number}{1} (\bibinfo{date}{jul}
  \bibinfo{year}{2018}), \bibinfo{pages}{1–96}.
\newblock
\showISSN{1935-8237}
\urldef\tempurl%
\url{https://doi.org/10.1561/2200000070}
\showDOI{\tempurl}


\bibitem[\protect\citeauthoryear{Russo, Hashimoto, Kang, Sun, and
  Zaharia}{Russo et~al\mbox{.}}{2022}]%
        {techreport}
\bibfield{author}{\bibinfo{person}{Matthew Russo}, \bibinfo{person}{Tatsunori
  Hashimoto}, \bibinfo{person}{Daniel Kang}, \bibinfo{person}{Yi Sun}, {and}
  \bibinfo{person}{Matei Zaharia}.} \bibinfo{year}{2022}\natexlab{}.
\newblock \bibinfo{title}{InQuest Technical Report}.
\newblock
\newblock
\urldef\tempurl%
\url{https://github.com/stanford-futuredata/InQuest/blob/main/inquest_technical_report.pdf}
\showURL{%
Retrieved Dec. 30, 2022 from \tempurl}


\bibitem[\protect\citeauthoryear{Stats}{Stats}{2013}]%
        {twitter2013data}
\bibfield{author}{\bibinfo{person}{Internet~Live Stats}.}
  \bibinfo{year}{2013}\natexlab{}.
\newblock \bibinfo{title}{Twitter Usage Statistics}.
\newblock
\newblock
\urldef\tempurl%
\url{https://www.internetlivestats.com/twitter-statistics/}
\showURL{%
Retrieved Nov. 27, 2022 from \tempurl}


\bibitem[\protect\citeauthoryear{Tracker}{Tracker}{2022}]%
        {twitch2022data}
\bibfield{author}{\bibinfo{person}{Twitch Tracker}.}
  \bibinfo{year}{2022}\natexlab{}.
\newblock \bibinfo{title}{Twitch broadcast time for all channels by month}.
\newblock
\newblock
\urldef\tempurl%
\url{https://twitchtracker.com/statistics/stream-time}
\showURL{%
Retrieved Nov. 27, 2022 from \tempurl}


\bibitem[\protect\citeauthoryear{Xarchakos and Koudas}{Xarchakos and
  Koudas}{2021}]%
        {xarchakos2021querying}
\bibfield{author}{\bibinfo{person}{Ioannis Xarchakos} {and}
  \bibinfo{person}{Nick Koudas}.} \bibinfo{year}{2021}\natexlab{}.
\newblock \showarticletitle{Querying for Interactions}.
\newblock \bibinfo{journal}{\emph{IEEE Transactions on Knowledge and Data
  Engineering}} (\bibinfo{year}{2021}), \bibinfo{pages}{1--1}.
\newblock
\urldef\tempurl%
\url{https://doi.org/10.1109/TKDE.2021.3094997}
\showDOI{\tempurl}


\bibitem[\protect\citeauthoryear{Zaharia, Das, Li, Hunter, Shenker, and
  Stoica}{Zaharia et~al\mbox{.}}{2013}]%
        {zaharia2013spark}
\bibfield{author}{\bibinfo{person}{Matei Zaharia}, \bibinfo{person}{Tathagata
  Das}, \bibinfo{person}{Haoyuan Li}, \bibinfo{person}{Timothy Hunter},
  \bibinfo{person}{Scott Shenker}, {and} \bibinfo{person}{Ion Stoica}.}
  \bibinfo{year}{2013}\natexlab{}.
\newblock \showarticletitle{Discretized Streams: Fault-Tolerant Streaming
  Computation at Scale}. In \bibinfo{booktitle}{\emph{Proceedings of the
  Twenty-Fourth ACM Symposium on Operating Systems Principles}} (Farminton,
  Pennsylvania) \emph{(\bibinfo{series}{SOSP '13})}.
  \bibinfo{publisher}{Association for Computing Machinery},
  \bibinfo{address}{New York, NY, USA}, \bibinfo{pages}{423–438}.
\newblock
\showISBNx{9781450323888}
\urldef\tempurl%
\url{https://doi.org/10.1145/2517349.2522737}
\showDOI{\tempurl}


\bibitem[\protect\citeauthoryear{Zhang, Ananthanarayanan, Bodik, Philipose,
  Bahl, and Freedman}{Zhang et~al\mbox{.}}{2017}]%
        {zhang2017live}
\bibfield{author}{\bibinfo{person}{Haoyu Zhang}, \bibinfo{person}{Ganesh
  Ananthanarayanan}, \bibinfo{person}{Peter Bodik}, \bibinfo{person}{Matthai
  Philipose}, \bibinfo{person}{Paramvir Bahl}, {and}
  \bibinfo{person}{Michael~J. Freedman}.} \bibinfo{year}{2017}\natexlab{}.
\newblock \showarticletitle{Live Video Analytics at Scale with Approximation
  and {Delay-Tolerance}}. In \bibinfo{booktitle}{\emph{14th USENIX Symposium on
  Networked Systems Design and Implementation (NSDI 17)}}.
  \bibinfo{publisher}{USENIX Association}, \bibinfo{address}{Boston, MA},
  \bibinfo{pages}{377--392}.
\newblock
\showISBNx{978-1-931971-37-9}
\urldef\tempurl%
\url{https://www.usenix.org/conference/nsdi17/technical-sessions/presentation/zhang}
\showURL{%
\tempurl}


\bibitem[\protect\citeauthoryear{Zhang and Kumar}{Zhang and Kumar}{2020}]%
        {zhang2020panorama}
\bibfield{author}{\bibinfo{person}{Yuhao Zhang} {and} \bibinfo{person}{Arun
  Kumar}.} \bibinfo{year}{2020}\natexlab{}.
\newblock \showarticletitle{Panorama: A Data System for Unbounded Vocabulary
  Querying over Video}.
\newblock \bibinfo{journal}{\emph{Proc. VLDB Endow.}} \bibinfo{volume}{13},
  \bibinfo{number}{4} (\bibinfo{date}{jan} \bibinfo{year}{2020}),
  \bibinfo{pages}{477–491}.
\newblock
\showISSN{2150-8097}
\urldef\tempurl%
\url{https://doi.org/10.14778/3372716.3372721}
\showDOI{\tempurl}


\end{thebibliography}
